\documentclass[twocolumn, prx, superscriptaddress,reprint]{revtex4-2}
\usepackage{graphicx}
\usepackage[usenames,dvipsnames]{color}
\usepackage{ulem}
\usepackage{CJK}
\usepackage[colorlinks, linkcolor=blue,anchorcolor=blue,citecolor=blue,urlcolor=blue]{hyperref}
\usepackage{physics}
\usepackage{comment}
\usepackage{titlesec}

\newcommand{\sx}{\sigma_x}

\newcommand{\sz}{\sigma_z}

\usepackage{color,soul}

\begin{document}
\begin{CJK*}{UTF8}{}
\title{Manipulating solid-state spin concentration through charge transport}
\author{Guoqing Wang \CJKfamily{gbsn}(王国庆)}\email[]{gq\_wang@mit.edu}
\affiliation{
   Research Laboratory of Electronics, Massachusetts Institute of Technology, Cambridge, MA 02139, USA}
\affiliation{
   Department of Nuclear Science and Engineering, Massachusetts Institute of Technology, Cambridge, MA 02139, USA}
 
\author{Changhao Li} 
\thanks{Present address: Global Technology Applied Research, JPMorgan Chase, New York, NY 10017 USA}
\affiliation{
   Research Laboratory of Electronics, Massachusetts Institute of Technology, Cambridge, MA 02139, USA}
\affiliation{
   Department of Nuclear Science and Engineering, Massachusetts Institute of Technology, Cambridge, MA 02139, USA}
   
\author{Hao Tang}
\affiliation{Department of Materials Science and Engineering, Massachusetts Institute of Technology, MA 02139, USA}   

\author{Boning Li}
\affiliation{
   Research Laboratory of Electronics, Massachusetts Institute of Technology, Cambridge, MA 02139, USA}
\affiliation{Department of Physics, Massachusetts Institute of Technology, Cambridge, MA 02139, USA}

\author{Francesca Madonini}
\affiliation{
   Research Laboratory of Electronics, Massachusetts Institute of Technology, Cambridge, MA 02139, USA}
\affiliation{
   Dipartimento di Elettronica, Informazione e Bioingegneria (DEIB), Politecnico di Milano, Piazza Leonardo da Vinci 32, 20133, Milano, Italy}

\author{Faisal F Alsallom} 
\affiliation{Department of Physics, Massachusetts Institute of Technology, Cambridge, MA 02139, USA}

\author{Won Kyu Calvin Sun} 
\affiliation{
   Research Laboratory of Electronics, Massachusetts Institute of Technology, Cambridge, MA 02139, USA}

\author{Pai Peng} 
\affiliation{
   Department of Electrical Engineering, Princeton University, Princeton, NJ 08544, USA}

\author{Federica Villa}
\affiliation{
   Dipartimento di Elettronica, Informazione e Bioingegneria (DEIB), Politecnico di Milano, Piazza Leonardo da Vinci 32, 20133, Milano, Italy}

\author{Ju Li}\email[]{liju@mit.edu}
\affiliation{
   Department of Nuclear Science and Engineering, Massachusetts Institute of Technology, Cambridge, MA 02139, USA}
\affiliation{Department of Materials Science and Engineering, Massachusetts Institute of Technology, MA 02139, USA}   

\author{Paola Cappellaro}\email[]{pcappell@mit.edu}
\affiliation{
   Research Laboratory of Electronics, Massachusetts Institute of Technology, Cambridge, MA 02139, USA}
\affiliation{
   Department of Nuclear Science and Engineering, Massachusetts Institute of Technology, Cambridge, MA 02139, USA}
\affiliation{Department of Physics, Massachusetts Institute of Technology, Cambridge, MA 02139, USA}

\begin{abstract}
Solid-state spin defects are attractive candidates for developing quantum sensors and simulators. The spin and charge degrees of freedom in large defect ensemble are a promising platform to explore complex many-body dynamics and the emergence of quantum hydrodynamics~\cite{zu_emergent_2021}. However, many interesting properties can be revealed only upon changes in the density of defects, which instead is usually fixed in material systems. Increasing the interaction strength by creating denser defect ensembles also brings more decoherence. Ideally one would like to control the spin concentration at will, while keeping fixed decoherence effects. Here we show that by exploiting charge transport, we can take some first steps in this direction, while at the same time characterizing charge transport and its capture by defects. By exploiting the cycling process of ionization and recombination of NV centers in diamonds, we pump electrons from the valence band to the conduction band. These charges are then transported to modulate the spin concentration by changing the charge state of material defects. By developing a wide-field imaging setup integrated with a fast single photon detector array, we achieve a direct and efficient characterization of the charge redistribution process by measuring the complete spectrum of the spin bath with micrometer-scale spatial resolution. We demonstrate the concentration increase of the dominant spin defects by a factor of 2 while keeping the $T_2$ of the NV center, which also provides a potential experimental demonstration of the suppression of spin flip-flops via hyperfine interactions. Our work paves the way to studying many-body dynamics with temporally and spatially tunable interaction strengths in hybrid charge-spin systems.
\end{abstract}

\maketitle

\end{CJK*}

\begin{figure*}[htbp]
\centering \includegraphics[width=\textwidth]{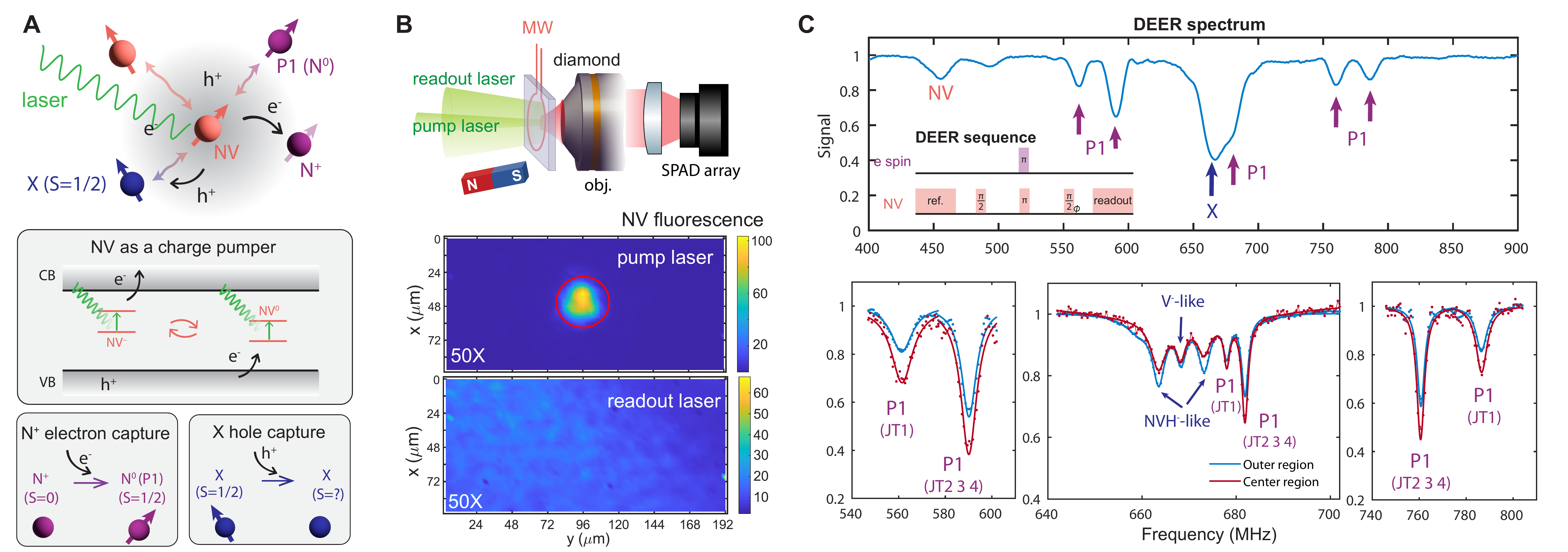}
\caption{\label{Fig_1} \textbf{Tuning the spin density via charge variations.} (A) Charge state conversion under laser illumination in a system containing different spin defects (top). The charge dynamics is influenced by laser illumination and electron/hole captures by different species (bottom) (B) Schematic of the experimental setup (top) and corresponding NV fluorescence imaging (bottom) under the illumination of the focused laser beam and the broad laser beam. (C) DEER spectra  when both  pump and readout lasers are on for 0.4~ms in each repetition (see Fig.~\ref{Fig_2}). The upper panel shows the measurement under a strong $\pi$ pulse power ($t_\pi=0.074\mu$s, calibrated for the 560~MHz P1 resonance frequency). The bottom three panels show the measurement under a selective $\pi$ pulse with low power (calibrated separately for each resonance frequency). Dark red curves show the total SPAD signal within the red circle shown in B, while the blue curves show the SPAD signal outside the circle, indicating different concentrations of the spin defects.
}
\end{figure*}


\section{Introduction}
Defects in solid state material have become promising quantum information platforms in developing quantum sensors~\cite{maze_nanoscale_2008,taylor_high-sensitivity_2008}, memories~\cite{ruskuc_nuclear_2022}, network~\cite{Pompili2022}, and simulators~\cite{zu_emergent_2021,wang_observation_2021-2}. In addition to the spin degree of freedoms that are most often used due to their valuable quantum coherence and controllability, the charge degree of freedom attracts increasing interest due to its influence on the spin state and potential in characterizing and tuning spin and charge environment~\cite{dolde_nanoscale_2014,dhomkar_charge_2018}. The charge state can be manipulated and probed through either optical illumination~\cite{aslam_photo-induced_2013,dhomkar_-demand_2018} or electrical gates~\cite{grotz_charge_2012,lozovoi_probing_2020,karaveli_modulation_2016}, and it can also serve as a meter to yield  information about the environment~\cite{DNAsensorACS2021,AFM2012,VoltageSensingNPhoton2022,IonSensor2301.03143}. Combining these control and detection methods,  charge transport  has been observed in diamond with both sparse~\cite{lozovoi_optical_2021} and dense~\cite{jayakumar_optical_2016} color centers. Moreover, non-fluorescent dark charge emitters can be imaged through carrier-to-photon conversion~\cite{lozovoi_dark_2020,lozovoi_imaging_2022}, and the spin-to-charge conversion has enabled single-shot spin readout~\cite{jayakumar_spin_2018,shields_efficient_2015,zhang_high-fidelity_2021}.

In this work, we show that charge transport can be used to manipulate the spin concentration in materials. Under optical illumination, the in-gap material defects undergoing cycling transitions of ionization and recombination can continuously pump electrons from the valence band (VB) to the conduction band (CB). Then the electrons diffuse and get captured by other in-gap defects, whose variation in charge state  modulates their spin states, thus affecting both  spin and charge environment simultaneously. With a home-built wide-field imaging microscope integrated with a fast single photon detector array and a two-beam pump-probe setup, we observe the charge pumping and redistribution among different spin defects. By characterizing the double electron-electron resonance (DEER) spectrum using NV centers in diamond, we show that the concentration of the dominant paramagnetic (S=1/2) nitrogen defect, the P1 center, can be increased by a factor of 2, while an additional, unknown-type electron spin density can be decreased by a factor of 2. 

\section{Tuning the spin density through charge transport} 
The sample we study  is a diamond grown using chemical-vapor-deposition (CVD) by Element Six with $\sim$10 ppm $^{14}$N concentration and 99.999\% purified $^{12}$C. Various in-gap point defects including nitrogen-vacancy (NV) centers are generated through electron irradiation and subsequent annealing at high temperature. 
The energy gap of a diamond is 5.5 eV, prohibiting pumping electrons from the VB to CB through a one-photon process with visible light sources around 2 eV. Instead, such a pumping can be assisted by these in-gap defects including the NV centers~\cite{jayakumar_optical_2016,lozovoi_optical_2021}. Under laser illumination with sufficient energy, the ionization of the negatively charged NV$^-$ is a two-photon process: it is first pumped to its excited state and is further ionized to NV$^0$ by transferring one electron to the diamond CB. The recombination of NV$^0$ is through a similar two-photon process with the second step pumping an electron from the valence band to convert the center back to NV$^-$~\cite{aslam_photo-induced_2013}. Such a cycling process generates a stream of electrons and holes in the conduction and valence bands respectively as shown in Fig.~\ref{Fig_1}A, which will subsequently transport diffusively to locations a few micrometers away in a time scale from milliseconds to seconds~\cite{jayakumar_optical_2016,lozovoi_optical_2021}.

In addition to photo-ionization, the conversion between different charge states can happen through direct electron or hole capture. Accounting for all these effects,  the densities of the negative NV charge states, $Q_-$, e.g., satisfy the equation    $\frac{dQ_-}{dt}=-k_-Q_-+k_0Q_0+\kappa_nnQ_0-\kappa_ppQ_-$, where $Q_0$ is the neutral NV density, $k_-,k_0$ are photo-ionization rate, $\kappa_n,\kappa_p$ are the electron trapping (releasing) rates, and $n,p$ are densities of electrons and holes. Similar equations apply to other defects with the corresponding rate constants set by the laser intensity and wavelength, as well as the charge state energies with respect to the CB and VB. The dynamic of the defect density distribution is further determined by including the modified diffusion equation for the electron and holes, e.g., $\frac{dn}{dt}=D_n(\partial^2n/\partial r^2+(1/r)\partial n/\partial r)+(e^-\text{ pumping}-e^-\text{capture})$ where the last two terms  describe the total rates of electron generation (pumping) and absorption (capture)  in the material, for example due to photo-(de)ionization or direct electron/hole capture process. 

When considering an (quasi-)equilibrium condition with (quasi-)static charge state density, the charge state distribution are set by the local electron and hole density, as well as the photo-ionization rate, which can be controlled through photo-ionization process. In this work, we focus on studying the substitutional nitrogen defects N$_s^0$ and N$_s^+$, where the neutral charge state N$_s^0$ (P1 center) has a spin $S=1/2$ and the positive charge state N$_s^+$ has no spin. In a typical CVD diamond, the density of these defects are usually one to two orders of magnitude larger than NV centers, thus contributing to the dominant spin bath~\cite{peaker_first_nodate}. 

Following the rate model above, the spin density can be obtained from $n[\text{N}_s^0]=n[\text{N}_s]\frac{\gamma_n n}{\gamma_n n+\gamma_p p +k_N}$, where constant $\gamma_n,\gamma_p$ are electron (hole) capture rates and $k_N$ is the P1 photoionization rate (only  N$_s^0$ $\rightarrow$  N$_s^+$ is considered due to insufficient laser energy for the inverse process in our experiment~\cite{jayakumar_optical_2016,pan_carrier_1990}). 
Then, the spin density $n[N_s^0]$ can be tuned through both the photoionization rate $k_N$ and the local charge densities $n,p$, in turn controlled by the NV charge cycling process and by charge transport. Previous studies have reported that the $n[$N$_s^+]$ in CVD diamond is typically 1-10 times more abundant than $n[$N$_s^0]$~\cite{edmonds_production_2012}, indicating a potentially large tunable range for $n[$N$_s^0]$.

\begin{figure*}[htbp]
\centering \includegraphics[width=0.99\textwidth]{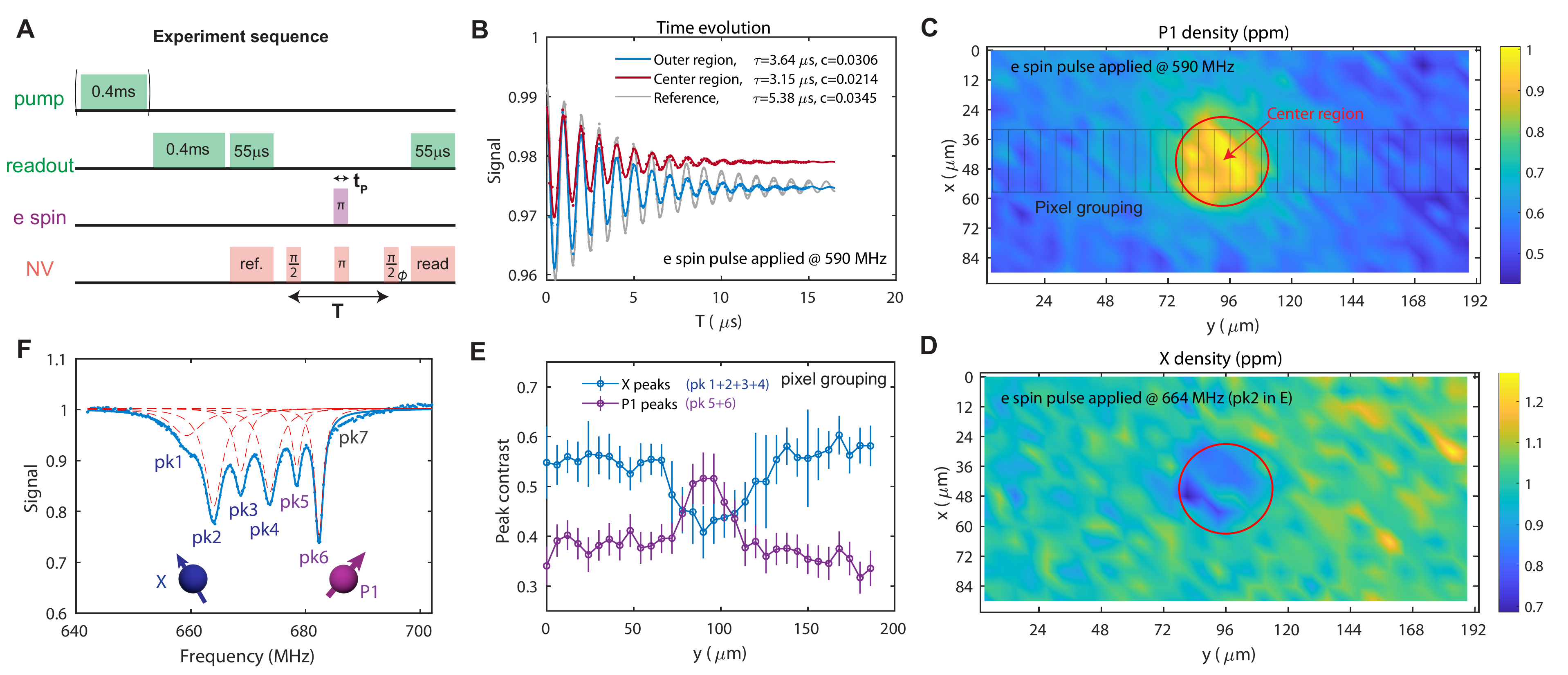}
\caption{\label{Fig_2} \textbf{Redistribution of electronic spin species under illumination.} (A) Experimental sequence used for the DEER experiment . (B) Time evolution of the NV signal under DEER in different regions of the sample. The  red curve is the total signal inside  the circle shown in C, while the blue curve is from outside the circle. The gray curve is the signal without applying the bath $\pi$ pulse (a typical spin echo). All the time-dependent signal measurements are fit to a decay exponential curve $S(T)=c_0+\frac12 c e^{-T(1/T_2+1/T_s^*)}\cos(d\omega T+\phi_0)$, where $\phi=d\omega T$ is a time-dependent phase applied  to the last NV $\pi/2$ pulse to modulate the signal. We set $d\omega=(2\pi)1~$MHz in all similar experiments in this work. The coherence times $T_s^*$ are used to determine the recoupled spin density. (C,D) imaging of P1 and X density (obtained with a DEER at 590 and 664~MHz resonance frequencies)  extracted from the measured coherence time $T_s^*$ according to Eq.~\eqref{eq:Ts*}.  (E) The X and P1 spectra peak height as a function of the $y$ position. The pixel signal is first grouped as shown by the gray rectangles in C for a better signal-to-noise ratio and then fit to six Lorentzian peaks to extract the peak heights as shown in (F) Representative DEER spectrum with high frequency resolution. The measured signal is fit to six Lorentzian peaks shown by the red dashed lines. 
}
\end{figure*}

\begin{figure}[htbp]
\centering \includegraphics[width=0.49\textwidth]{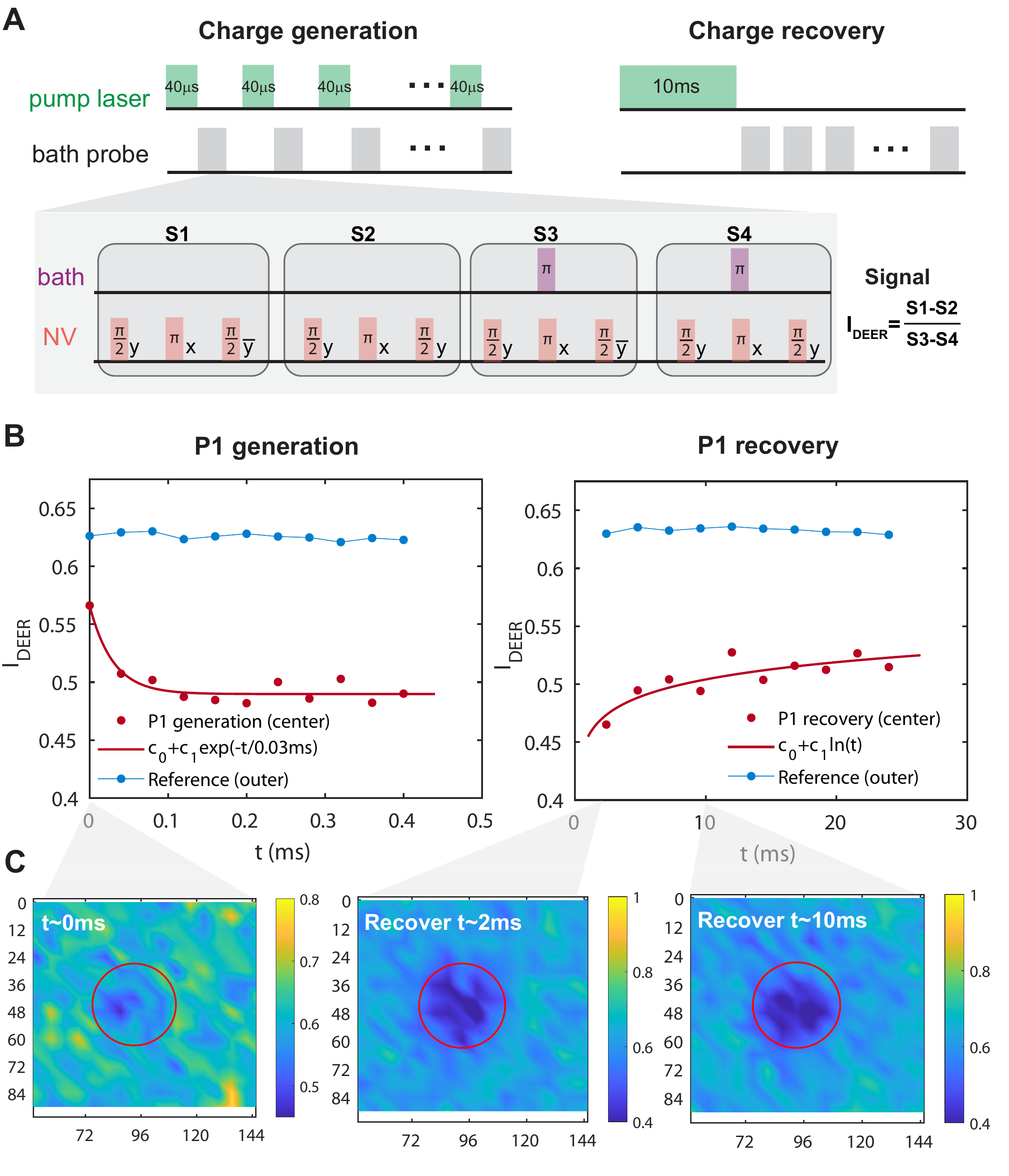}
\caption{\label{Fig_3} \textbf{Lifetime of charge states.} (A) Experimental sequence to probe the charge dynamics. To measure the generation of charges via their effect on P1 density, we apply a  short 0.04~ms pump laser pulses to better capture the fast generation process. The laser is followed by an echo or DEER sequence that measures the P1 concentration (DEER) and calibrate it against the NV coherence (echo). The echo/DEER are further calibrated by repeating the sequence with a different phase of the last $\pi$/2 pulse. A 0.126~$\mu$s $\pi$ pulse is applied to probe the P1 on resonance at 590~MHz. To measure the charge recovery, a single long laser illumination of 10ms duration (to ensure sufficient charge generation and transport) is followed by repeated echo/DEER sequences. (B). Evolution of the DEER contrast $I_{\textrm{DEER}}$ in center and outer regions during and after the charge pumping.   (C). The contrast $I_{\textrm{DEER}}$ imaging at various recovery time durations. 
}
\end{figure}

\section{Probing spin environment with DEER}
To probe the electronic spin environment with high resolution over both spatial and frequency domains, we combine our novel imaging setup with DEER control techniques. 

We  built a wide-field imaging setup~\cite{wang_fast_2023} integrated with a fast single photon avalanche diode (SPAD) array~\cite{madonini_single_2021} with a 100 kHz frame rate and $32\times 64$ pixels~\cite{bronzi_100_2014}, as shown in Fig.~\ref{Fig_1}B. A narrow laser beam ($D\sim32\mu$m, 0.5 W) is used for charge pumping whereas a broad laser beam ($D\sim200-500\mu$m, 0.6 W) is used for fluorescence readout. A static magnetic field $B=239~$G is aligned along one of the four NV orientations. To  probe the various spin species interacting with our NV ensemble, we apply the DEER sequence (inset of Fig.~\ref{Fig_1}C) to perform spectroscopy of the spin bath. As shown in Fig.~\ref{Fig_1}C, the spectrum shows that, in addition to the expected P1 centers, a substantial concentration of extra (Ex) defects exist with resonance near $g=2$. 

To study the effect of our controlled charge pumping process (via controlled NV illumination), we analyze the spin concentration quantitatively from the DEER measurements. In the DEER experiment (Fig.~\ref{Fig_2}A), the NV acts as a probe of the magnetic noise generated by the component of the spin bath that is resonantly driven. More specifically, while the spin echo sequence (alone) applied on the NV cancels out the interaction with the entire electronic spin bath, the addition of a spin flip pulse close to a particular bath-spin resonance recouples the interaction with only the corresponding spin species. The measured DEER signal is then composed of a Ramsey-like decoherence characterized by ${T_s^*}$, due to interaction with the recoupled spin species, and an echo-like decay due to the other spin species, with characteristic time $\sim T_2$, with a form $I_{\textrm{DEER}}=\exp{-T(\frac{1}{T_s^*}+\frac{1}{T_2})}$, where $T$ is the free evolution time.
The dephasing time due to a given spin species can be calculated from the rms noise field acting  on the NV center due to the spin dipolar coupling ~\cite{li_determination_2021,stepanov_determination_2016}, yielding  
\begin{equation}
    \frac{1}{T_s^*}=\frac{2\pi\mu_0\mu_B^2g_Ag_s|\sigma|}{9\sqrt{3}\hbar}P_s n_s = 0.1455n_s [\mu s^{-1}]
    \label{eq:Ts*}
\end{equation}
where $g_A,g_s$ are $g$-factors for the central and bath spins, $n_s$ is the concentration of the spin species $s$ in ppm.  $P_s$ describes the probability to invert the spin population, and we set   $P_s=1$ assuming the $\pi$ pulse for spin flip operation is perfect, and bath spins are assumed to be spin-1/2,  thus $|\sigma|=1/2$.
Here  $T_s^*$ is in units of $\mu$s. 

To extract the spin density, we experimentally measure the NV coherence time under the DEER sequence for each resonance frequency and fit the exponential decay to the analytical formula, $I_{\textrm{DEER}}$. By comparing the experiments with and without the pump beam shown in Fig.~\ref{Fig_2}A, we extract the value of $T_s^*$, from which we obtain the density of the recoupled spin species $n_s$. In Fig.~\ref{Fig_2}B we show an example of the comparison of time evolution experiments with (blue) and without (gray) recoupling pulse, and with pump laser illumination (red). The measurements show clear differences in both the coherence times and signal contrast, revealing the different bath spin density and NV ionization fraction (see Supplemental Materials~\cite{SOM} for more details).

\begin{figure*}[htbp]
\centering \includegraphics[width=0.99\textwidth]{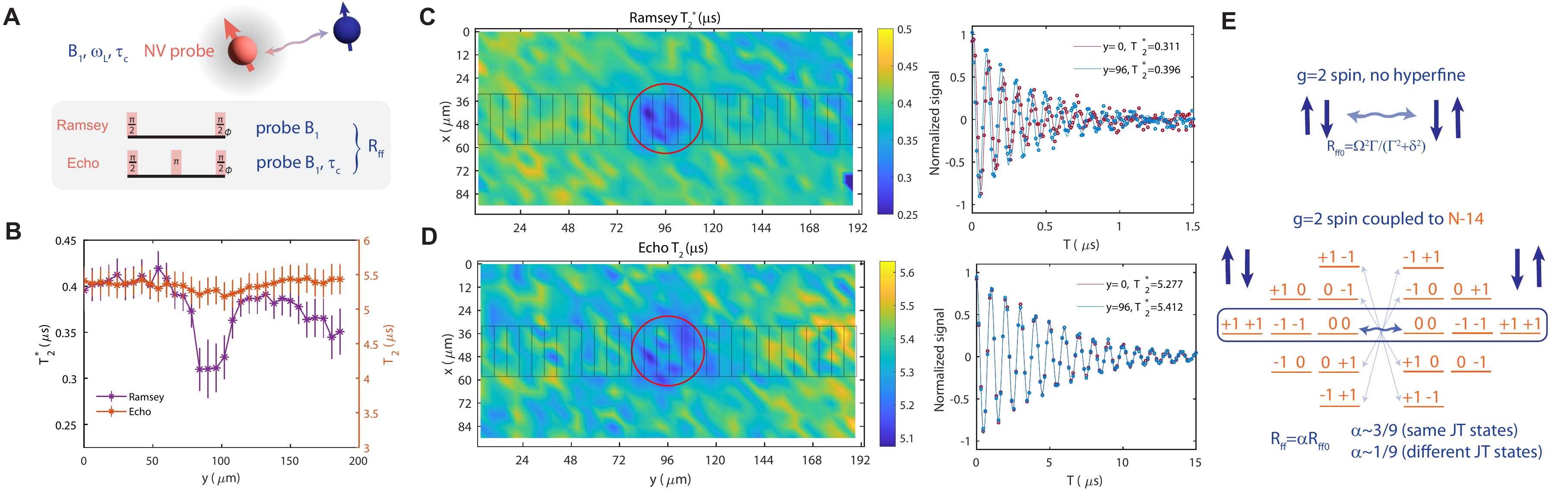}
\caption{\label{Fig_4} \textbf{Coherence time imaging.} (A). Ramsey and echo sequences to probe the spin density and flip-flop rates of the spin bath surrounding the NV probe. (B). Coherence time under Ramsey and echo experiments. The complete sequence including lasers is  similar to Fig.~\ref{Fig_2}A. Each data point is the fitted decay time of the total signal in grouped pixels shown C,D. (C). Ramsey coherence time $T_2^*$ imaging. A large coherence time decrease is observed in the pump beam region (see the range of the color bar). (D). Echo coherence time $T_2$ imaging. A small coherence time decrease is observed in the center region. (E). Analytical model for the suppression of electronic spin flip-flop rate.
}
\end{figure*}

\section{Redistribution of electronic spin species under laser illumination}
Applying the above techniques to probe our sample with moderate native spin densities and sufficiently coherent NV ensemble, we report a high-resolution DEER spectrum around $g=2$ which reveals, for the first time, a (reproducible) redistribution of the electronic spin species in diamond under laser illumination. Concretely, a low-power DEER spectrum reveals multiple resonances (labeled Peak 1-6 in Fig.~\ref{Fig_2}F), which includes not only the known P1 (Peak 5-6) but also additional resonances (Peak 1-4) belonging to spins dubbed X defects in the literature~\cite{li_determination_2021}.  Thus comparing the DEER spectra with and without the strong focused green laser reveals redistribution of spin species with a preferential increase  for P1 and decrease for X, shown in Figs.~\ref{Fig_2}C,D,E.

Controlled variations in the spin bath density are a useful resource for quantum simulation of many-body physics, as recently proposed~\cite{zu_emergent_2021}. 
Concretely, to further enhance the quantum simulation capability of NV-P1 systems, we would like to add: (i) a tunable knob over [P1] that is also stable for the duration of a typical quantum simulation ($\mu$s-ms), which however (ii) leaves intact the coherence properties of the NV centers. The former capability increases the range of interaction strengths that can be explored in the quantum simulator, while the latter will at least keep the same circuit depth/simulation time despite the cost of enacting (i) which physically leads to a non-equilibrium complex charge environment.

We now show that we can take steps towards realizing these two goals  via the NV charge cycling and transport process. 

\subsection{Tunable and stable P1 concentration in the dark}
In order to be useful for quantum simulations, we should be able to change the density of spin defects with illumination, but then maintain it at a stable level over the spin dynamics characteristic timescale (from $\mu$s to ms given  typical dipolar interaction strength of MHz to kHz). To verify that  the charge state can be treated as quasi-static thanks to its long   recovery timescale in the dark (previously evaluated to be due on the ms to seconds timescale) ~\cite{lozovoi_optical_2021,jayakumar_optical_2016}, 
we experimentally characterize the charge dynamics including both the generation and recovery processes using the sequence shown in Fig.~\ref{Fig_3}A. After a fast photo-induced charge initialization process (characteristic timescale less than 0.05~ms), the charge state shows a long recovery time,  more than tens of ms, as shown in Figs.~\ref{Fig_3}B,C. Our results demonstrate the feasibility and flexibility of manipulating the spin concentration using charge transport in a quasi-static way. We note that the measurement of the recovery shape and rate is a characterization of the underlining material properties including charge mobility, capture cross-section, tunneling rate, etc. Further quantitative explanation of such a process requires a more comprehensive study combining different experimental probes and theoretical insights.

\subsection{Coherence time under varying spin density}
The coherence times of NV centers are influenced by the redistribution of the spin density.
 
The coherence time of  a central single spin (NV) interacting with a dipolar spin bath  can be estimated by assuming the bath  generates a classically fluctuating 
 field~\cite{bauch_decoherence_2020}. The central spin coherence time under free evolution (Ramsey dephasing) is then $T_2^*=\frac{\sqrt{2}}{\gamma_e\sqrt{\langle B_1^2\rangle}}$, while evolution  under a spin echo lead to a coherence time $T_2=\left(\frac{12\tau_c}{\gamma_e^2\langle B_1^2\rangle}\right)^{1/3}.$
The mean strength of the noise field $\sqrt{\langle B_1^2\rangle}$ is proportional to the spin density, and the correlation time $\tau_c$ is set by the \textit{flip-flop} interaction between resonant bath spins. Combining the analysis of $T_2$ and $T_2^*$ provides a characterization of both the density and the correlation of the spin bath~\cite{sun_self-consistent_nodate} as shown in Fig.~\ref{Fig_4}A. In Figs.~\ref{Fig_4}B-D, we measure the spatial distribution of both $T_2^*$ and $T_2$ with the pump laser illumination applied to the center region. The experimental results show that even though  $T_2^*$  decreases in the center region due to the increase of total spin densities,  $T_2$ remains relatively unchanged for all regions.

One potential explanation of such an intriguing phenomena is based on the  suppression of flip-flop rates of different spin species due to hyperfine interactions and Jahn-Teller (JT) orientations, which introduce energy mismatch in the two-spin flip-flop subspace. Such an effect was discussed in Refs.~\cite{bauch_decoherence_2020,wang_spin_2013} and was recently revealed by first-principles in Ref.~\cite{park_decoherence_2022}. In the flip-flop subspace spanned by the states $\ket{0}=\ket{\uparrow\downarrow}$ and $\ket{1}=\ket{\downarrow\uparrow}$,  the interaction Hamiltonian is simplified to $H_{sub}=\delta\sz+C_\perp\sx$
with $2\delta$ the energy difference between states $\ket{0}$ and $\ket{1}$ and $C_\perp$ the transverse coupling strength. 
Taking into account the dephasing rate $\Gamma_d$, the bath correlation time $\tau_c$ is then given by $1/{\tau_c}\sim R_{ff,0}^{ij}={(C_\perp^{ij})^2\Gamma_d}/{(\Gamma_d^2+\delta^2)}$~\cite{bauch_decoherence_2020}.
The different nuclear spin states and JT orientations $\ket{m}$, suppress the flip-flop rate  as they induce the energy mismatch $\delta_m$ between $\ket{\uparrow\downarrow}\otimes \ket{m_1}$ and $\ket{\downarrow\uparrow}\otimes\ket{m_2}$ ($\delta_m\gg\Gamma_d$). Intuitively, we can define a suppression factor $\alpha$, in comparison to the un-suppressed flip-flop rate $R_{ff,0}$, by summing over all cases of $\ket{m}$ such that $R_{\textrm{ff}}^{ij}=\frac{(C_\perp^{ij})^2}{\Gamma_d}\sum_m \frac{1}{M} \frac{\Gamma_d^2}{(\Gamma_d^2+\delta_m^2)}=\alpha R_{ff,0}^{ij}$.
$M$ is the number of nuclear spin states and JT orientations as shown in Fig.~\ref{Fig_4}E~\footnote{Here we assume the nuclear spin is in a fully-mixed state such that different states $\ket{m}$ contribute in an equal probability. The magnetic field is aligned along JT1 orientation such that JT2,3,4 orientations are degenerate.}.
Numerical calculations give the suppression factors for different spin defects $\alpha_{P1}\approx0.208$, $\alpha_{NVH}\approx0.353$ assuming $\Gamma_d=(2\pi)1~$MHz (see SM~\cite{SOM} for different approaches to calculating $\alpha$).

With large hyperfine interaction, the electronic spin flip-flop of P1 is more suppressed than other defects, especially those appearing near $g=2$. Thus, the contribution to the change of $T_2$ time due to density decrease of NVH-like spin defects ($\sim$2.5~ppm to $\sim$1.6~ppm) could compensate the larger amount of P1 increase ($\sim$2~ppm to $\sim$4~ppm). Our experiments present an observation of different suppressions of the noise contribution as predicted by the recent work~\cite{park_decoherence_2022}. We note that the measured $T_2^*$ in Fig.~\ref{Fig_4}C is shorter than what we predict from the total spin density ($T_2^*$ is predicted to be $\sim$1.1~$\mu$s for the outer region given by the characterized spin densities). We attribute such a discrepancy to the magnetic or electric field inhomogeneity generated by the charge transport, which creates local current induced by the laser on-off during the sequence repetitions and electron-hole separation due to their different diffusion constants~\cite{SOM}.

\section{Discussion on the identification of X}

Even though we cannot uniquely identify the X spins, we can still draw a few conclusions based on experiments and theoretical calculations. Peaks 1,2,4 in Fig.~\ref{Fig_1}C and Fig.~\ref{Fig_2}F show a larger contrast decrease in the center region than the peak 3 at the $g=2$ frequency. The widths and contrasts of peaks 2, 4 are similar, and both  are larger than peak 3. Based on this, we conclude that we have at least two spin species, peaks 1,2,4 belonging to X1 and peak 3 belonging to X2. Comparing the spectrum features with simulations shown in Supplementary Materials~\cite{SOM} based on reported values of various spin defects in CVD diamond~\cite{peaker_first_nodate}, X1 is possibly the NVH$^{-}$ defect and X2 can be any $g=2$ spin defect without hyperfine splitting, such as a simple vacancy V$^-$ (or a spin with a weak hyperfine such as VH$_2^-$). Different from the increase of P1 due to the electron capture of the N$_s^+$, the decrease of X1 and X2 can be induced by the hole capture process or direct photoionization. For NVH$^{-}$, the hole capture process converts it to a spin-1 state NVH$^{0}$. Using first-principles calculation, we obtain the zero-field splitting of NVH$^{0}$ as $D=2.290$ GHz, which predicts the spin resonance frequencies. However, no significant signals are observed at these frequencies, indicating the potential further ionization (if exists)~\cite{SOM}. Indeed, the laser energy 532~nm=2.33~eV used in our experiment is larger than the first-principle predicted excitation (1.603~eV) and ionization (2.27~eV) energies for NVH$^{0}$. Further identification of these defects can be achieved by varying excitation laser frequencies to characterize these transition energies.

\section{Discussions and conclusions}
We propose and demonstrate the manipulation of spin concentration in diamonds through charge transport. Making use of the redistribution of the photo-induced charges over different spin species with different flip-flop suppression, the coherence time of the probe (NV) spin is preserved. Besides characterizing the steady states of the system under repetition of the sequence, we demonstrate the feasibility of characterizing the charge dynamics. Our work provides a flexible tool to characterize materials,  including both the charge and spin dynamics by measuring the DEER spectrum.

In recent years, solid state defects have shown potential in exploring hydrodynamics, aiming to bridging the gap between microscopic quantum laws and macroscopic classical phenomena~\cite{zu_emergent_2021}. The natural dipolar interactions in the spin ensemble serves as a versatile platform to engineer and characterize many-body quantum spin systems~\cite{zu_emergent_2021,davis_probing_2021}. Our work provides approaches to temporally and spatially tune the spin concentration in these spin transport experiments in the same material while maintaining the spin coherence time. Moreover, introducing the charge degree of freedom into the system provides a more flexible platform to engineer a many-body system with two coupled transport mechanisms~\cite{zu_emergent_2021,doherty_towards_2016,ku_imaging_2020}.
For example, with the capability of reading out an array of pixels in a fast and parallel manner, our experiment provides a platform to study the spin transport described by $\partial_tP(t,\Vec{r})=D(t,\Vec{r})\nabla^2 P(t,\Vec{r})-P(t,\Vec{r})/T_1+Q(t,\Vec{r})$ with a temporally and spatially varying diffusion constant $D(t,\Vec{r})$.

Further integrating our setup with a wavelength-tunable laser could reveal the defect energy levels with respect to the materials energy bands more precisely, allowing  identification of various defects when combining experiments with first-principles calculations. Such a fingerprint enables a more complete reconstruction of the local charge and spin environment~\cite{lozovoi_probing_2020,rezai_probing_2022,zhang_reporter-spin-assisted_2022}. 
Combining the charge density control with spin-to-charge conversion~\cite{irber_robust_2021} paves the way to coherent transport of quantum information through charge carriers~\cite{doherty_towards_2016} and on-demand generation of other quantum spin defects~\cite{zhang_neutral_2022}. Besides, improving the optical tunability of local charge density is promising for developing charge lense through laser beam patterning~\cite{sivan_electrostatic_1990}.

\section*{Methods}
\subsection{Ab-initio calculations of the vacancy defect spin Hamiltonian}
We implemented  ab-initio  calculations of the NVH$^0$ and VH$_2^0$ defects in diamond. The defect structure is simulated by the supercell method~\cite{nieminen2007supercell} using a $3\times 3\times 3$ cubic supercell (with 216 atoms). The calculation used density functional theory (DFT) and projector-augmented wave (PAW) method carried out through Vienna $ab-initio$ simulation package (VASP)~\cite{kresse1996efficient,kresse1999ultrasoft}. Generalized gradient approximation is used for exchange-correlation interaction with the PBE functional~\cite{perdew1996generalized}. The cut-off energy is set as 400 eV, and a $3\times 3\times 3$ $k$-point mesh is sampled by Monkhorst-Pack scheme~\cite{monkhorst1976special}. The defect structure is fully relaxed with a residue force on each atom less than 0.01 eV/\AA , and the electronic energy converges to $10^{-7}$ eV. The zero-field splitting is then calculated from the DFT electronic ground state using~\cite{ivady2014pressure}:
\begin{equation*}
D_{a b}=\mu_0 g_e^2 \beta^2 \sum_{i<j}^{p+q} \chi_{i j}\langle\Psi_{i j}(\mathbf{r}_1, \mathbf{r}_2)|\frac{r^2 \delta_{a b}-3 r_a r_b}{r^5}| \Psi_{i j}(\mathbf{r}_1, \mathbf{r}_2)\rangle.
\end{equation*}
The calculated $D_{ab}$ of NVH$^0$ is:
\begin{equation*}
D_{a b}(NVH^0)=\left[\begin{matrix}-0.76 &0 &0 \\ 0& -0.76& 0 \\ 0& 0& 1.53\end{matrix}\right]  \text{GHz},
\end{equation*}
where the $z$-axis is set as the $C_{3v}$ axis of NVH$^0$. The calculated $D_{ab}$ of VH$_2^0$ in the cubic axis (the axis is along the three lattice vectors of the cubic supercell of diamond, and we make the two hydrogen atoms in $xz$-plane) is:
\begin{equation*}
D_{a b}(VH_2^0)=\left[\begin{matrix} -0.40 &0 & 2.21 \\ 0& 0.79& 0 \\ 2.21 & 0& -0.40\end{matrix}\right]  \text{GHz},
\end{equation*}

\section*{Acknowledgements}
It is a pleasure to thank Carlos A. Meriles, Artur Lozovoi, Tom Delord, Richard Monge, and Bingtian Ye for useful discussions. This work was partly supported by DARPA DRINQS program (Cooperative Agreement No. D18AC00024). F.M. acknowledges the Rocca program for support and Micro Photon Devices S.r.l. for providing the MPD-SPC3 camera. 

\section*{Author contributions}
G.W., C.L., and P.C. proposed the method and designed the experiment. G.W. performed the experiment and data analysis, with partial contributions from B.L., F.M., W.K.C.S., and P.P. H.T. performed first-principles calculations of the spin transition frequencies. P.C. and J.L. supervised the study. G.W. wrote the paper with contributions from all authors.

\section*{Competing financial interest}
The authors declare no competing financial interests.
\bibliography{main_text}


\newpage
\clearpage

\widetext
\setcounter{section}{0}
\setcounter{equation}{0}
\setcounter{figure}{0}
\setcounter{table}{0}
\setcounter{page}{1}
\makeatletter
\renewcommand{\theequation}{S\arabic{equation}}
\renewcommand{\thefigure}{S\arabic{figure}}
\renewcommand{\thesection}{S\arabic{section}}


\begin{CJK*}{UTF8}{}
\title{Supplementary Materials: Manipulating solid-state spin concentration through charge transport}
\author{Guoqing Wang \CJKfamily{gbsn}(王国庆)}\email[]{gq\_wang@mit.edu}
\affiliation{
   Research Laboratory of Electronics, Massachusetts Institute of Technology, Cambridge, MA 02139, USA}
\affiliation{
   Department of Nuclear Science and Engineering, Massachusetts Institute of Technology, Cambridge, MA 02139, USA}
 
\author{Changhao Li}
\thanks{Current address: Global Technology Applied Research, JPMorgan Chase, New York, NY 10017 USA}
\affiliation{
   Research Laboratory of Electronics, Massachusetts Institute of Technology, Cambridge, MA 02139, USA}
\affiliation{
   Department of Nuclear Science and Engineering, Massachusetts Institute of Technology, Cambridge, MA 02139, USA}
   
\author{Hao Tang}
\affiliation{Department of Materials Science and Engineering, Massachusetts Institute of Technology, MA 02139, USA}   

\author{Boning Li}
\affiliation{
   Research Laboratory of Electronics, Massachusetts Institute of Technology, Cambridge, MA 02139, USA}
\affiliation{Department of Physics, Massachusetts Institute of Technology, Cambridge, MA 02139, USA}

\author{Francesca Madonini}
\affiliation{
   Research Laboratory of Electronics, Massachusetts Institute of Technology, Cambridge, MA 02139, USA}
\affiliation{
   Dipartimento di Elettronica, Informazione e Bioingegneria (DEIB), Politecnico di Milano, Piazza Leonardo da Vinci 32, 20133, Milano, Italy}

\author{Faisal F Alsallom} 
\affiliation{Department of Physics, Massachusetts Institute of Technology, Cambridge, MA 02139, USA}

\author{Won Kyu Calvin Sun} 
\affiliation{
   Research Laboratory of Electronics, Massachusetts Institute of Technology, Cambridge, MA 02139, USA}

\author{Pai Peng} 
\affiliation{
   Department of Electrical Engineering, Princeton University, Princeton, NJ 08544, USA}

\author{Ju Li}\email[]{liju@mit.edu}
\affiliation{
   Department of Nuclear Science and Engineering, Massachusetts Institute of Technology, Cambridge, MA 02139, USA}
\affiliation{Department of Materials Science and Engineering, Massachusetts Institute of Technology, MA 02139, USA}   

\author{Paola Cappellaro}\email[]{pcappell@mit.edu}
\affiliation{
   Research Laboratory of Electronics, Massachusetts Institute of Technology, Cambridge, MA 02139, USA}
\affiliation{
   Department of Nuclear Science and Engineering, Massachusetts Institute of Technology, Cambridge, MA 02139, USA}
\affiliation{Department of Physics, Massachusetts Institute of Technology, Cambridge, MA 02139, USA}
\onecolumngrid
\maketitle
\end{CJK*}

\newpage
\onecolumngrid
\tableofcontents

\section{Derivation of coherence times}
\subsection{Signal decay due to an effective stochastic classical environment}
For both Ramsey and Echo experiments, the measured signal can be expressed as
\begin{equation}
    S(T) = \frac{1}{2}[1+\langle \cos(\varphi(T))\rangle],
\end{equation}
where the phase $\varphi$ is
\begin{equation}
    \varphi(T)=\gamma_e\int dtf(t)B_1(t).
\end{equation}
Here $B_1(t)$ is the effective classical random field generated by the spin bath acting on the sensor, $f(t)$ is the modulation function describing the external $\pi$-pulse control on the NV spin sensors. In most experimental situations with large number of bath spins coupled to the NV center, we can assume a Gaussian distribution for the phase $\phi$, and the signal can be further written as
\begin{equation}
    S(T)=\frac12(1+e^{-\chi})=\frac12(1+e^{-\langle\varphi^2\rangle/2}),
\end{equation}
with $\chi$ the second cumulant of the noise
\begin{equation}
    \chi=\frac12\langle\varphi^2\rangle=\frac{\gamma_e^2}2\int_0^T\int_0^T dt_1 dt_2 \langle B_1(t_1)B_1(t_2)\rangle f(t_1)f(t_2).
    \label{Supp_eq:chi_theory}
\end{equation}
We define the auto-correlation of the noise as $G(t_1,t_2)$ 
\begin{equation}
    G(t_1,t_2)=\gamma_e^2\langle B_1(t_1)B_1(t_2)\rangle.
    \label{Supp_eq:G_theory}
\end{equation}
For stationary and zero-mean noise, the correlation only depends on the difference between two times,  that is, $G(t_1,t_2)=G(|t_1-t_2|)$.

The coherence decay rate $\chi$ in (\ref{Supp_eq:chi_theory}) can be also expressed as an integral in frequency space
\begin{align}
    \chi&=\frac12\int_0^T\int_0^T dt_1 dt_2 f(t_1)f(t_2) \int_{-\infty}^{+\infty}\frac{d\omega}{2\pi}S(\omega)e^{i\omega (t_1-t_2)}\\&=\frac12\int_{-\infty}^{+\infty}\frac{d\omega}{2\pi}S(\omega)|F(\omega)|^2\label{Supp_eq:Filter}
\end{align}
where in (\ref{Supp_eq:Filter}) we defined the filter function $F(\omega)=|\int_0^Tf(t)e^{i\omega t}dt|$ associated with the pulse control sequence.

\subsection{Coherence time dependence on bath parameters}
The central spin decay times  $T_2$ and $T_2^*$ depend on the spin bath parameter and thus can reveal its characteristics. 
We start from analyzing a spin bath composed of same types of spins with the same energy (same Larmor frequency $\omega_L$~\cite{bauch_decoherence_2020}.) Due to dipolar interactions among the bath spins, the noisy field they produce $B(t)$ fluctuates in time, with a correlation time $\tau_c$  set by the strength of the transverse dipolar interaction, which induces spin flip-flops. The auto-correlation function can be modeled as $G(t)=\gamma_e^2 \langle B_1^2\rangle \cos(\omega_L t)e^{-t/\tau_c}$. Its associated spectrum is then Lorentzian, 
\begin{equation}
    S(\omega)=\frac{\gamma_e^2B_1^2\tau_c}{1+(\omega-\omega_L)^2\tau_c^2}+\frac{\gamma_e^2B_1^2\tau_c}{1+(\omega+\omega_L)^2\tau_c^2}.
\end{equation}

The filter function for Ramsey sequence is $|F(\omega)|^2=\frac{4\sin^2(\frac{\omega T}{2})}{\omega^2}$,
giving
\begin{align}
    \chi&=\frac1\pi \int_{-\infty}^{+\infty}\frac{\sin^2(\frac{\omega T}{2})}{\omega^2}S(\omega)d\omega\\
    &\approx \gamma_e^2B_1^2\tau_c^2\left[\frac{T}{\tau_c}-1+e^{-\frac{T}{\tau_c}}\right],
\end{align}
where the last line assumes $\omega_L=0$.
The filter function for the echo sequence is instead $|F(\omega)|^2=\frac{16\sin^4(\frac{\omega T}{4})}{\omega^2}$,
giving for $\omega_L=0$ 
\begin{equation}
    \chi\approx \gamma_e^2B_1^2\tau_c^2\left[\frac{T}{\tau_c}-3-e^{-\frac{T}{\tau_c}}+4e^{-\frac{T}{2\tau_c}}\right].
\end{equation}
When $t\ll\tau_c$ and $\gamma_eB_1\tau_c\gg 1$, we obtain 
\begin{equation}
    \chi_{Ramsey}=\left(\frac{T}{T_2^*}\right)^2,\qquad
    \chi_{Echo} = \left(\frac{T}{T_2}\right)^3
\end{equation}
with \begin{equation}
    T_2^*=\frac{\sqrt{2}}{\gamma_e\sqrt{\langle B_1^2\rangle}},\qquad T_2=\left(\frac{12\tau_c}{\gamma_e^2\langle B_1^2\rangle}\right)^{1/3}.
\end{equation}
In particular we see that while the dephasing time only depends on the rms noise strength $\sqrt{\langle B_1^2\rangle}$, the echo time (as well as  the coherence time under multi-$\pi$ sequences) is modulated by the noise correlation time. 

\subsection{Spin bath correlation time from first principle}
The spin bath density is expected to directly influence both the rms noise strength and its correlation time. 
However, for a complex electronic spin bath that is coupled to nuclear spins via hyperfine interactions, the bath flip-flop processes might be suppressed, and thus needs to  be calculated more accurately.


The dipolar interaction between two bath electronic  spins of the same kind $ij$ in the presence of a (large) magnetic field along the $z$ direction can be written as 
\begin{equation}
    H_{ij} = C(\theta)^{ij}[S_z^iS_z^j-\frac14(S_+^iS_-^j+S_-^iS_+^j)]
\end{equation}
where  $S_\pm=S_x\pm iS_y$ and only terms commuting with the Zeeman Hamiltonian are kept. The interaction strength is 
\begin{equation}
    C^{ij}(\theta)=\frac{3\cos(\theta_{ij})^2-1}{r_{ij}^3},
\end{equation}
where $\theta_{ij}$ is the angle between the vector $\vec r_{ij}$  and the $z$ direction.

The longitudinal term in the Hamiltonian conserves the total spin polarization, while the transverse term  induces spin flip-flops. For a pair of electronic spin-1/2, we can analyze such a spin flip-flop in a subspace spanned by the states $\ket{0}=\ket{\uparrow\downarrow}$ and $\ket{1}=\ket{\downarrow\uparrow}$, with corresponding effective Pauli operators $\sigma_\alpha$. Then, the interaction Hamiltonian can be further simplified to 
\begin{equation}
    H_{sub}=\delta\sz+C_\perp\sx
\end{equation}
where $2\delta$ is the energy difference between states $\ket{0}\ket{1}$ and $\ket{1}\ket{0}$ and $C^{ij}_\perp=C^{ij}(\theta)/4$. This Hamiltonian would give rise to (detuned) Rabi oscillations at the frequency $\sqrt{C_\perp^2+\delta^2}$. The energy difference $\delta$ is however not homogeneous, with a linewidth  $\Gamma_d\approx \sqrt{N_b}\Bar{C}_{||}^{ij}$ due to interactions to other spins. Thus the flip-flop rate between the two states is finally
\begin{equation}
R_{\textrm{ff}}^{ij}=\frac{(C_\perp^{ij})^2\Gamma_d}{\Gamma_d^2+\delta^2}.
\end{equation}
Similar to the idea proposed in Ref.~\cite{bauch_decoherence_2020}, we can define the spin bath correlation time $\tau_c$ to satisfy
\begin{equation}
    \frac{1}{\tau_c}=\sum_{ij}R_{\textrm{ff}}^{ij}=\sum_{ij}\frac{(C_\perp^{ij})^2\Gamma_d}{(\Gamma_d^2+\delta^2)}.
\end{equation}

When the bath electron spin has hyperfine interactions with nearby nuclear spins, the spin flip-flop is partly suppressed due to the energy mismatch between $\ket{\uparrow\downarrow}$ and $\ket{\downarrow\uparrow}$ ($\delta\gg\Gamma_d$)~\cite{park_decoherence_2022}. When the nuclear spins are fully mixed states, we can derive the suppression factor by summing over the flip-flop rate equation for all the $M$ nuclear spin sub-states $\ket{m}$ of both the electronic states $\ket{\uparrow\downarrow}\otimes \ket{m}$ and $\ket{\downarrow\uparrow}\otimes\ket{m}$
\begin{equation}
    R_{\textrm{ff}}^{ij}=\sum_m \frac{1}{M} \frac{(C_\perp^{ij})^2\Gamma_d}{(\Gamma_d^2+\delta_m^2)}=\alpha R_{\textrm{ff},0}^{ij},
\end{equation}
where we assume other parameters are fixed and only the energy detuning $\delta_m$ is different for different $m$. To simplify the analysis in the following discussions, we introduce a suppression factor $\alpha$ to compare the actual flip-flop rate to its maximum value $R_{\textrm{ff},0}^{ij}=(C_\perp^{ij})^2/\Gamma_d$. For free electrons without hyperfine interactions, the suppression factor is $\alpha=1$.

For  P1 centers in most existing diamonds (concentration of P1 smaller than 100~ppm), the hyperfine coupling strengths with nitrogen-14 nuclear spin,  $A_{zz}=(2\pi)114~$MHz, $A_{xx}=A_{yy}=-(2\pi)82~$MHz, are much larger than the dipolar-coupling induced linewidth $\Gamma_d\sim C^{ij}\lesssim$~MHz. The nuclear spin sub-states $\ket{m}$ have 9 possible combinations $\ket{++},\ket{+0},\ket{+-}\cdots$ as shown in Fig.~4E of the main text. 

We consider the case when one orientation of NV centers in the diamond is aligned to a static magnetic field $B$ with $\gamma_eB\gg A_{zz,xx,yy}$. The P1 centers have four Jahn-Teller (JT) orientations with different local $z$ axis to define their hyperfine interaction with the nuclear spin. We defined the aligned orientation to be JT1 and three misaligned degenerate orientations to be JT2,3,4. 
Since only the nuclear spin sub-states with the same  energy  contribute to the  electron spin flip-flop processes, we have $\alpha\approx\frac19$ for the aligned JT P1 spins, that can only interact with each other. Similarly, we find 
 $\alpha\approx 3/9$ for the flip-flop rate  between the degenerate JT orientations. 
 Summing over all the possibilities gives
\begin{equation}
    \alpha\approx\frac{3}{9}(\frac{1}{4^2}+\frac{3^2}{4^2})+\frac{1}{9}(2\times \frac{1}{4}\times\frac{3}{4})=\frac{1}{4}.
\end{equation}
Since $T_2\propto\tau_c^{1/3}\propto\alpha^{-1/3}$ while $T_2^*\propto\alpha^0$, comparing $T_2$ and $T_2^*$ makes it possible to analyze contributions from different bath spin species due to their different $\alpha$. 

In diamond samples grown by CVD methods, various other spin defects have been observed such as NVH$^-$ (S=1/2) or V$^{-}$ (S=3/2)~\cite{isoya_epr_1992,nunn_electron_2022,loubser_electron_1978,griffiths_paramagnetic_1954,baldwin_electron_1963}. As analyzed above, our sample have a significant concentration of spin defects near the $g=2$ frequencies including NVH$^-$-like and V$^{-}$-like defects. V$^{-}$-like defect can be treated as a parasitic electron spin 1/2 with $\alpha_V=1$. NVH$^-$ (S=1/2) defect has two nuclear spin with hyperfine interactions $A_{zz}(e-H)=(2\pi)13.69~$MHz, $A_{xx}(e-H)=-(2\pi)9.05$~MHz,  and $A_{zz}(e-N)=(2\pi)2.1~$MHz, $A_{xx}(e-N)=-(2\pi)2.2$~MHz~\cite{peaker_first_nodate}\footnote{The hyperfine interactions with nitrogen-14 nuclear spins are deduced from the reported experiment where nitrogen-15 were used by setting $A(e-^{14}{N})=A(e-^{15}{N})\gamma_{^{14}N}/\gamma_{^{15}N}$.}.

\begin{figure*}[htbp]
\centering \includegraphics[width=0.7\textwidth]{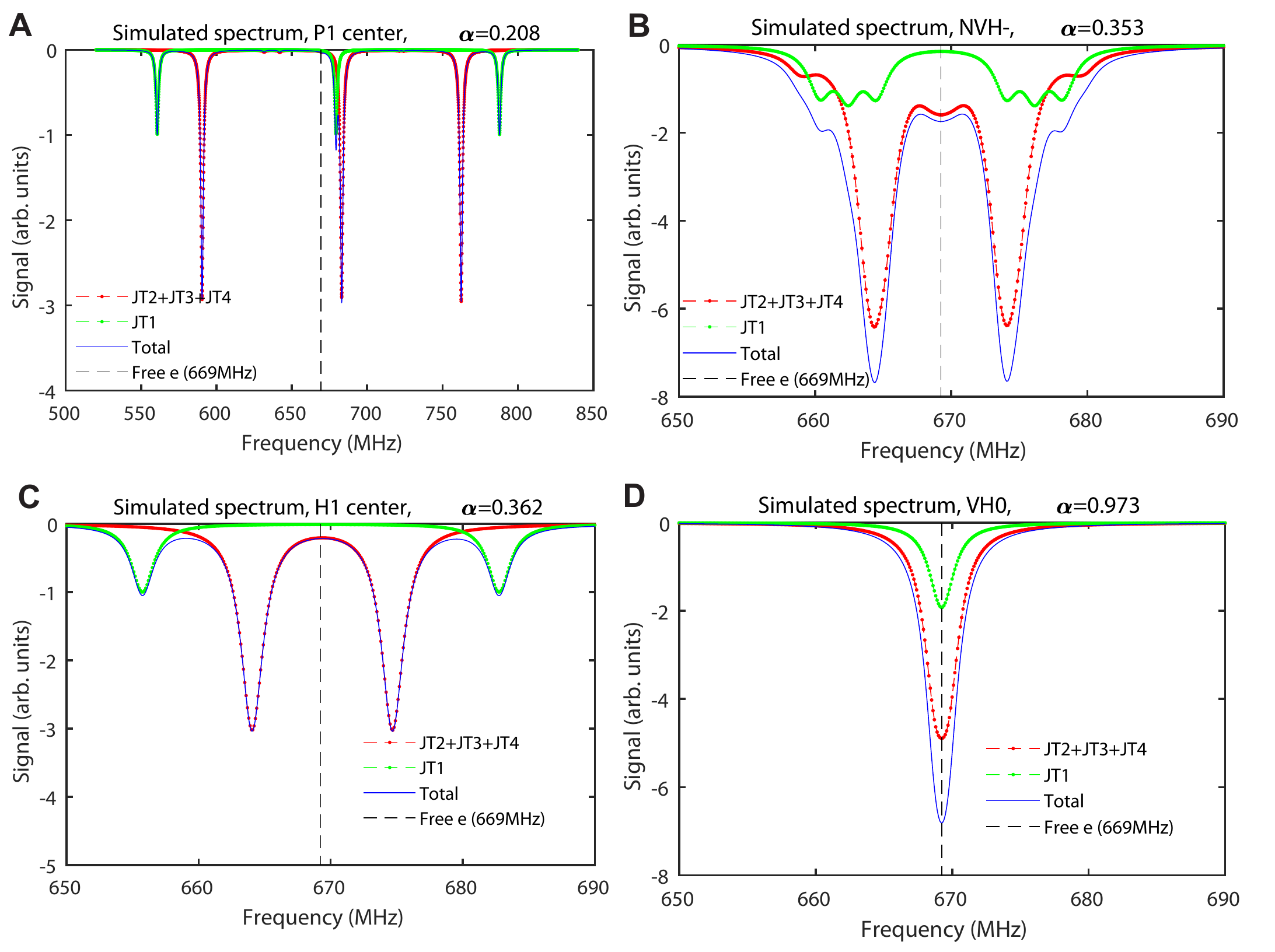}
\caption{\label{Fig_Spectrum} \textbf{Simulated spectrum.} (a) Simulated DEER spectrum for P1 center under a magnetic field $B=238.8~$G aligned to the JT1 state. A dephasing rate $\Gamma_d=(2\pi)1~$MHz is assumed. The same condition applies to the other simulations. (b) Simulated NVH$^{-}$ spectrum. (c) Simulated H1 center spectrum. (d) Simulated VH$^0$ spectrum. 
}
\end{figure*}

We note that above we assumed zero flip-flop between spins with different hyperfine states. However, due to the hyperfine mixing of states under a small-to-moderate external magnetic field (where the Zeeman energy is on the same order as the hyperfine), flip-flop between different JT orientations might be possible. We take this into account by performing exact diagonalization of the Hamiltonian, that reveal additional spin flip-flop transitions. 
To understand the $T_2$ under our experimental condition with a $B=239~$G magnetic field aligned to the diamond [111] axis, we calculate the values of $\alpha$ for different potential defects. Using exact diagonalization of the Hamiltonian of the system, we sum over all the spin flip-flop channels. Our calculation results are shown in Fig.~\ref{Fig_Spectrum}. Assuming $\Gamma_d=(2\pi)1~$MHz, our calculation gives $\alpha_{P1}=0.208$ (compared to 0.25 without exact diagonalization), $\alpha_{NVH}=0.353$, $\alpha_{H1}=0.362$, $\alpha_{VH0}=0.973$.

\subsection{Spin bath correlation time from spectra}
Instead of analyzing 9 different hyperfine levels with exact diagonalization of spin Hamiltonian, we can also directly look at the DEER spectrum and analyze the intensity of 6 different peaks. The overlapping peaks can be counted as a single peak where flip-flop happens. Such a method does not require a knowledge of the spin species, and can simplify the analysis and obtain a flip-flop suppression factor similar to the previous method.

The intensity or peak heights in the simulated DEER (or ESR) spectrum  reveal the probability distribution of different nuclear spin states. For example, in Fig.~\ref{Fig_Spectrum}(A) we observe six electronic spin  transition frequencies with intensity ratio $\frac{1}{12}:\frac{1}{4}:\frac{1}{12}:\frac{1}{4}:\frac{1}{4}:\frac{1}{12}$ when the nuclear spin is assumed to be a fully mixed state and different orientations of P1 centers have the same population. The spin flip-flop process, composed of two spin-flip transitions separately happening on two electronic spins, needs to conserve energy to have a large flip-flop rate. This implies that the flip-flop process mainly happens ``within" each peak, both of the flips of the two spins should have matched transition frequencies (their discrepancy should not be larger than the linewidth, similar to the condition $\delta\ll \Gamma_d$). Thus, comparing to the P1 centers without hyperfine interactions (all transitions are degenerate), the suppression factor $\alpha$ is then
\begin{equation}
    \alpha=\left[(\frac{1}{12})^2+(\frac{1}{4})^2\right]\times3=\frac{5}{24}\approx 0.208,
\end{equation}
which is exactly the same as what we derived by direct diagonalizing the Hamiltonian.

We note that due to the potential correlation of noise fields generated by different spin species under the ensemble average scenario, the practical flip-flop suppression factor might be more complicated. In addition to first-principles studies~\cite{park_decoherence_2022}, theoretical modeling of the flip-flop suppression factor (especially in spin ensembles) is of interest to future study.

\begin{figure*}[htbp]
\centering \includegraphics[width=\textwidth]{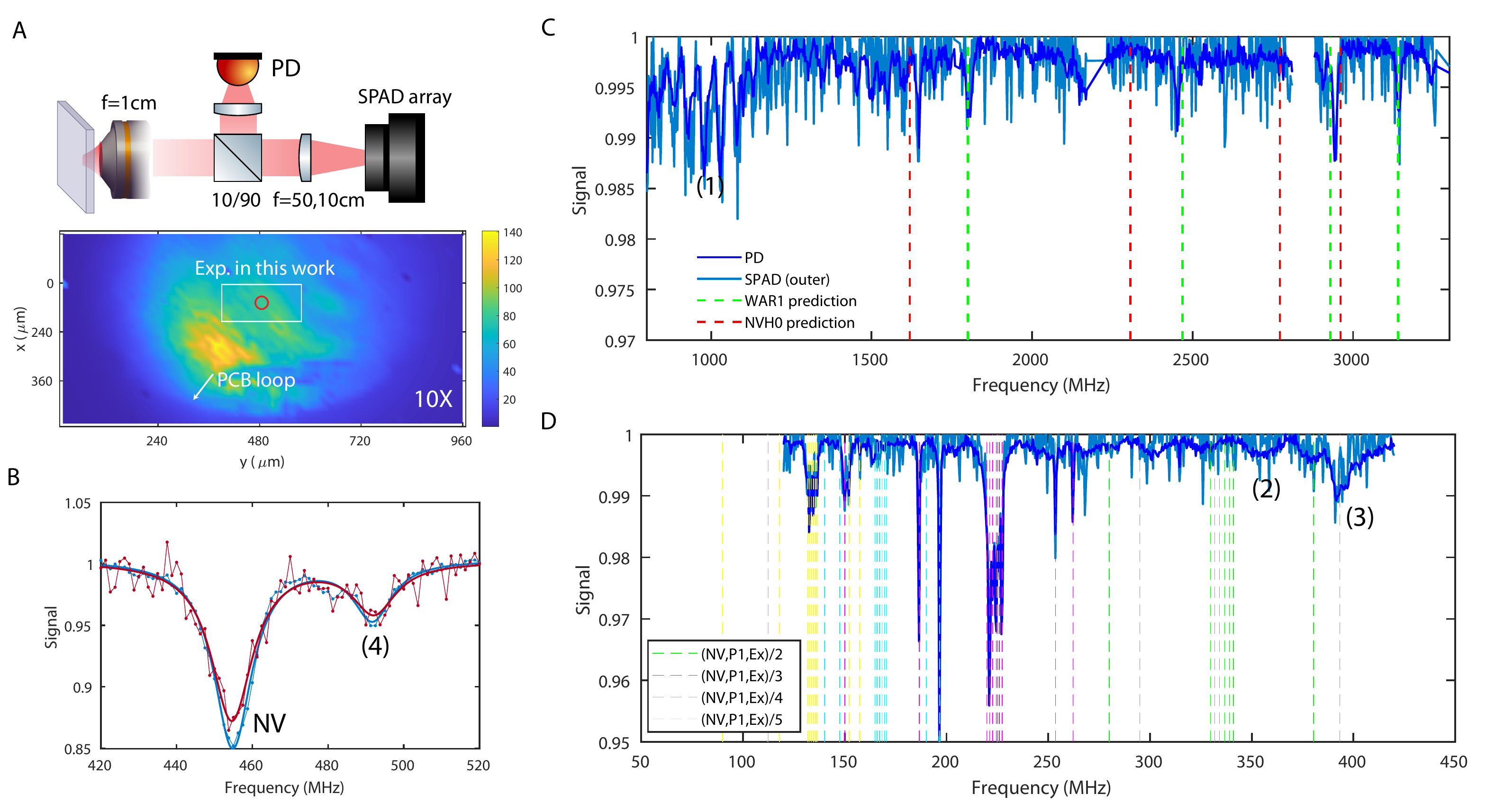}
\caption{\label{Supp_Fig_Spectrum} \textbf{DEER spectrum.} (A). Schematic of the optics and 2D mapping under the 10X magnification setting. The PCB loop structure and diamond surface is clearly seen in the mapping under 10X magnification. (B). DEER spectrum in the frequency range of 420~MHz to 520~MHz. The total signal measured in (out) the red circle is plotted in red (blue). (C,D). DEER spectrum in an extended frequency range. The signal measured by the photodiode is denoted as PD, while the signal measured by the SPAD camera is under 50X magnification setting.  
}
\end{figure*}

\section{Experiments}
\subsection{Setup}
The sample used in our experiment is a CVD-grown diamond from Element Six with $\sim$10 ppm $^{14}$N concentration and 99.999\% purified $^{12}$C. An  NA=0.50 objective (N20X-PF, Nikon Plan Fluorite) with 2.1 mm working distance and 10mm effective focal length is used to collect the fluorescence in a wide field. The collected photon is further split to two paths through a 10:90 beam splitter, with 10\% collected by the single photon avalanche detector (SPAD) array, and 90\% collected by a photodiode (PDA36A). Another achromatic lens with focal length 500~mm (100~mm) focuses the fluorescence beam to the SPAD camera, achieving a magnification factor of 50X (10X) as shown in Fig.~\ref{Supp_Fig_Spectrum}. High-pass filters with a cutoff wavelength 594~nm are used for both the SPAD array and the photodiode. We apply a static magnetic field $B=238.8$~G such that the transition frequency between NV$^-$ electronic spin states $\ket{m_s=0}$ and $\ket{m_s=-1}$ is 2.2~GHz. A 1~mm diameter loop fabricated on a 0.4~mm thin printed circuit board is used to deliver the microwave. The microwave to control the NV center is amplified by an amplifier with model ZHL-16W-43-S+, and the microwave to control the P1 center is amplified by an amplifier with model ZHL-20W-202-S+. Outputs of the two amplifiers are connected to two ports of the PCB. Other details on the microwave circuits are reported in Refs.~\cite{wang_characterizing_2022,wang_sensing_2022}.

\subsection{DEER spectrum}
In the main text, we focus on discussing the spectrum between 500~MHz to 800~MHz where the P1 and X spin defects have major contributions. To perform a more complete analysis of the existing spin defects, we also scan a much larger spectrum range as shown in Fig.~\ref{Supp_Fig_Spectrum}.

In the main text, we discuss the possibility of the X spins to be NVH$^-$. Under the illumination of the pump laser, the concentration of NVH$^-$ decreases, which might convert to neutrally charged NVH$^0$ which is a spin-1 defect. However, in the spectrum in Fig.~\ref{Supp_Fig_Spectrum}B, we do not see a clear signal at the predicted frequencies of NVH$^0$. Combining with the fact given by the first-principles calculation that the excitation and ionization energies of NVH$^0$ is smaller than the laser energy 2.33~eV, the  NVH$^0$ might be further excited or ionized and cannot be easily detected. Even though NVH$^0$ is not clearly observed, we found some signal near the resonance frequencies of WAR1 defect predicted by the zero field splitting reported in Ref.~\cite{pellet-mary_optical_2021}.

In Fig.~\ref{Supp_Fig_Spectrum}B, we see the double-quantum transition frequency of misaligned NV centers, which is forbidden for aligned ones. The contrast decrease in the center region indicates the NV density decrease, which is also measured from the signal contrast change shown in Figs.~\ref{SuppFig_1},\ref{SuppFig_2},\ref{SuppFig_3},\ref{SuppFig_4},\ref{SuppFig_5}.

In Fig.~\ref{Supp_Fig_Spectrum}C, we see some resonances at location (1) around 1000~MHz. Since they form a manifold consisting five to seven equally spacing peaks, we believe this could be some spin defects with multiple same nuclear spins ($^{14}$N). In addition, the observed peaks at locations (2), (3), (4) in Fig.~\ref{Supp_Fig_Spectrum}B,D are unknown. The identification of these peaks requires further works combining with other detection approaches. We note that due to the imperfection of electronics used in this work, we see signal when the higher harmonics of the microwave frequency matches the bath spin resonance frequencies as shown in Fig.~\ref{Supp_Fig_Spectrum}D.

\subsection{Details on more experiments}
In the main text, we reported the time evolution measurement under the DEER sequence recoupling the P1 resonance at 590~MHz or the X resonance at 664~MHz. Here we include more details about the complete measurement.

As discussed in the main text, the measured DEER signal is  due to a Ramsey-like dephasing, with rate $1/{T_s^*}$ induced by the recoupled spin species and an echo-like decay induced by the other spin species with rate $1/T_{2}^\prime$. The signal decay can thus be written as $I_{\textrm{DEER}}=\exp{-T(\frac{1}{T_s^*}+\frac{1}{T_{2}^\prime})}$ (assuming for simplicity they both have a simple exponential behavior). We note that  $T_{2}^\prime$ is different
from $T_2$ which includes contributions of all the spin species. A more rigorous derivation should in principle consider such a difference. However, since in practical DEER experiment we only recouple a small part of the total spin species $(<1/5)$, the analysis in the main text uses $T_2^\prime\approx T_2$.
Analyzing the average noise field on the NV center due to the dipolar coupling ~\cite{li_determination_2021,stepanov_determination_2016} further gives 
\begin{equation}
    \frac{1}{T_s^*}=\frac{2\pi\mu_0\mu_B^2g_Ag_B|\sigma|}{9\sqrt{3}\hbar}P_B n_s = 0.1455n_s.
    \label{Supp_eq:Ts*}
\end{equation}

To extract the spin concentration $n_s$, we measure the $I_{\textrm{DEER}}(T)$ by sweeping the evolution time $T$ and compare the coherence time with $T_2$ measured by the spin echo experiments. Then we extract the $n_s$ from Eq.~\eqref{Supp_eq:Ts*}. 

\begin{figure*}[htbp]
\centering \includegraphics[width=\textwidth]{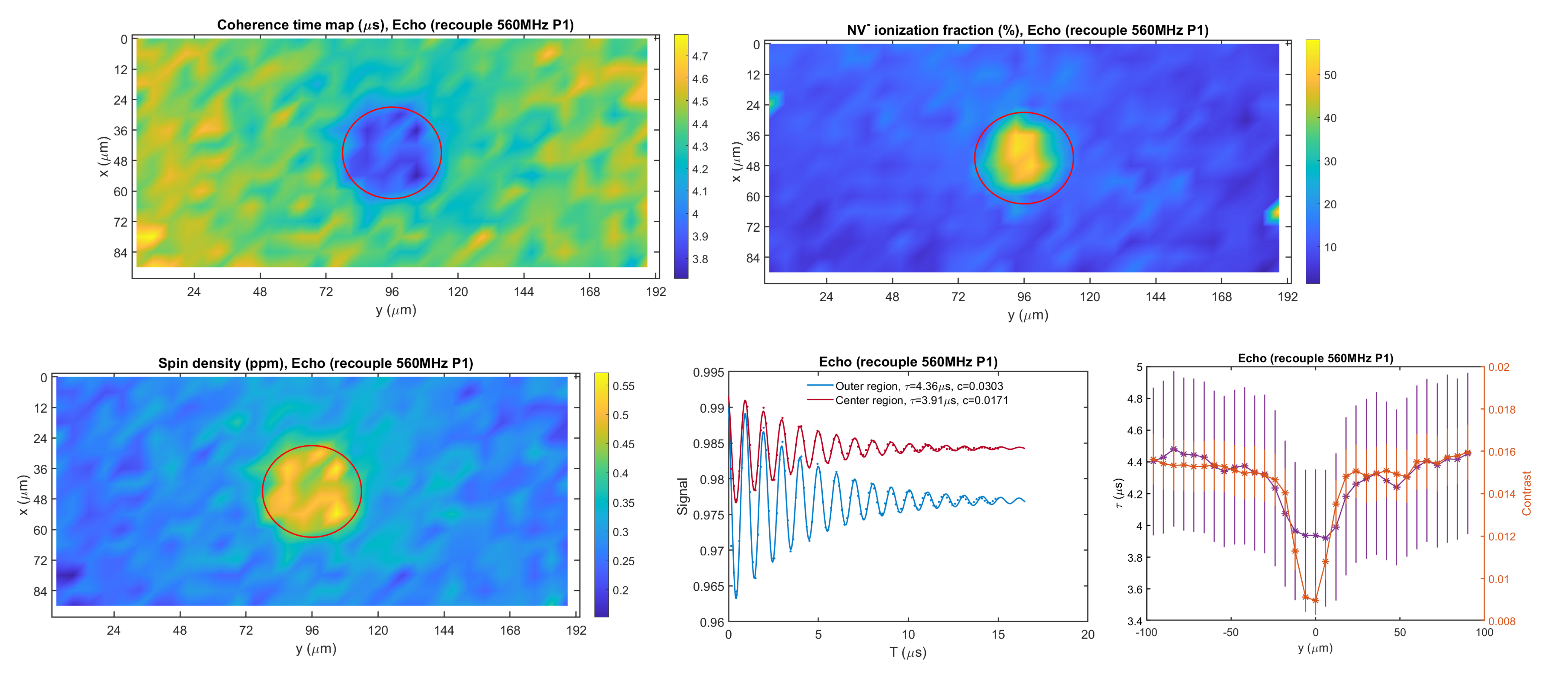}
\caption{\label{SuppFig_1} \textbf{DEER time evolution experiment with 560~MHz P1 recoupled.} 
}
\end{figure*}

\begin{figure*}[htbp]
\centering \includegraphics[width=\textwidth]{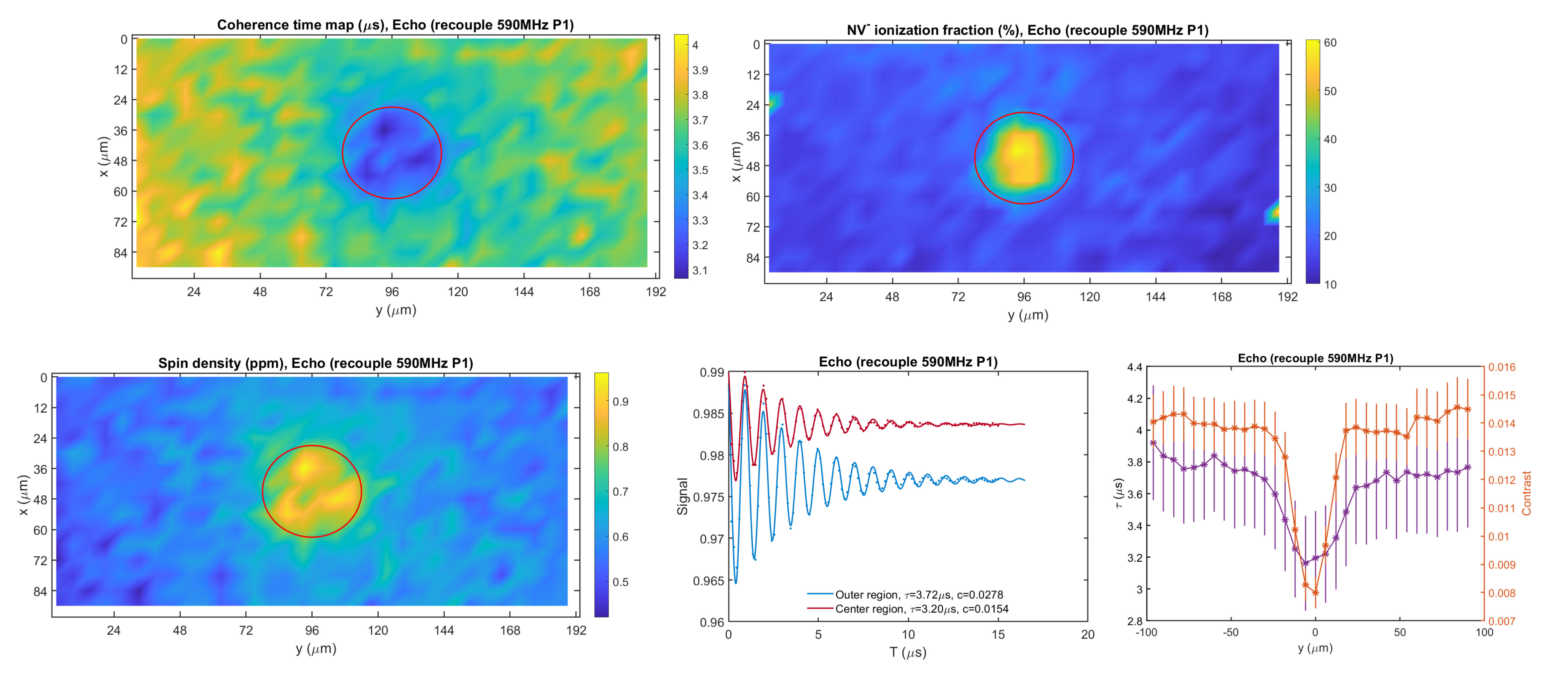}
\caption{\label{SuppFig_2} \textbf{DEER time evolution experiment with 590~MHz P1 recoupled.}
}
\end{figure*}

\begin{figure*}[htbp]
\centering \includegraphics[width=\textwidth]{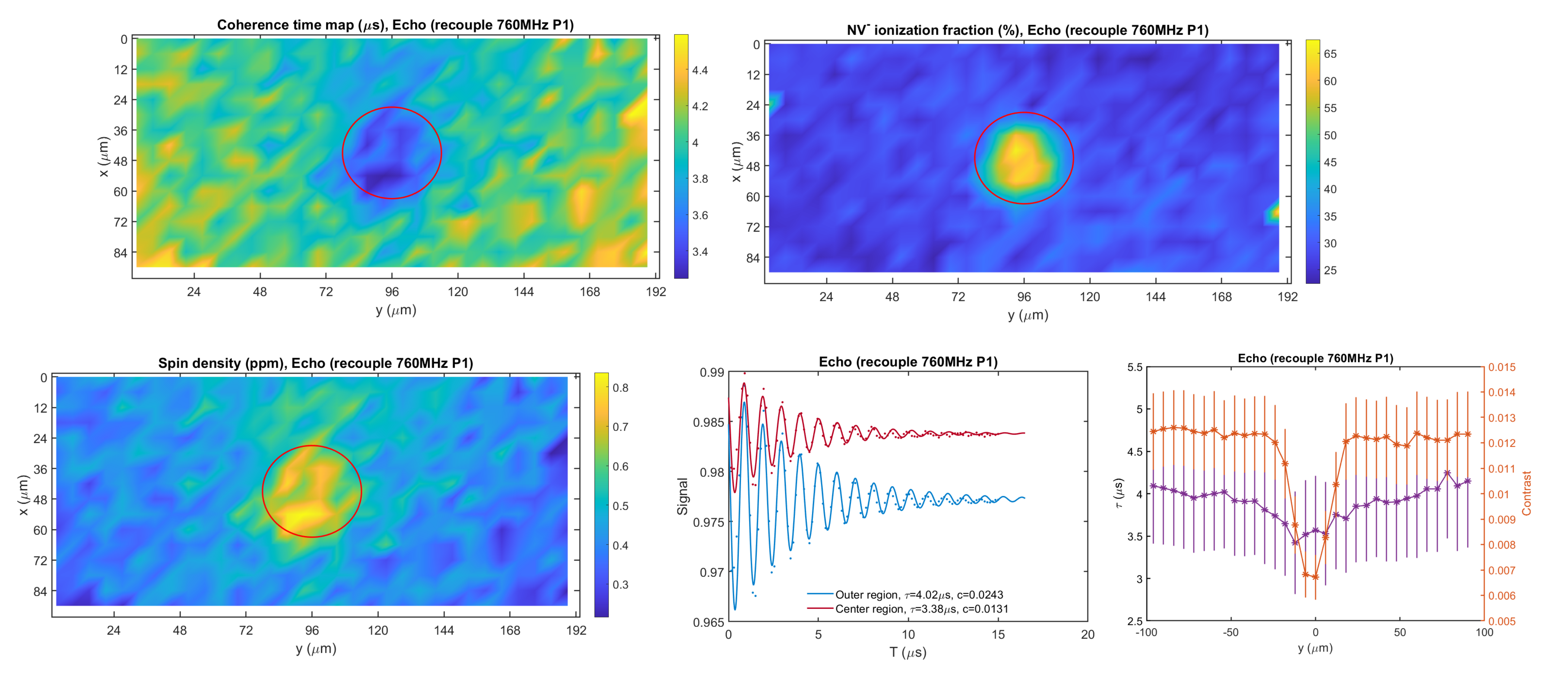}
\caption{\label{SuppFig_3} \textbf{DEER time evolution experiment with 760~MHz P1 recoupled.}
}
\end{figure*}

\begin{figure*}[htbp]
\centering \includegraphics[width=\textwidth]{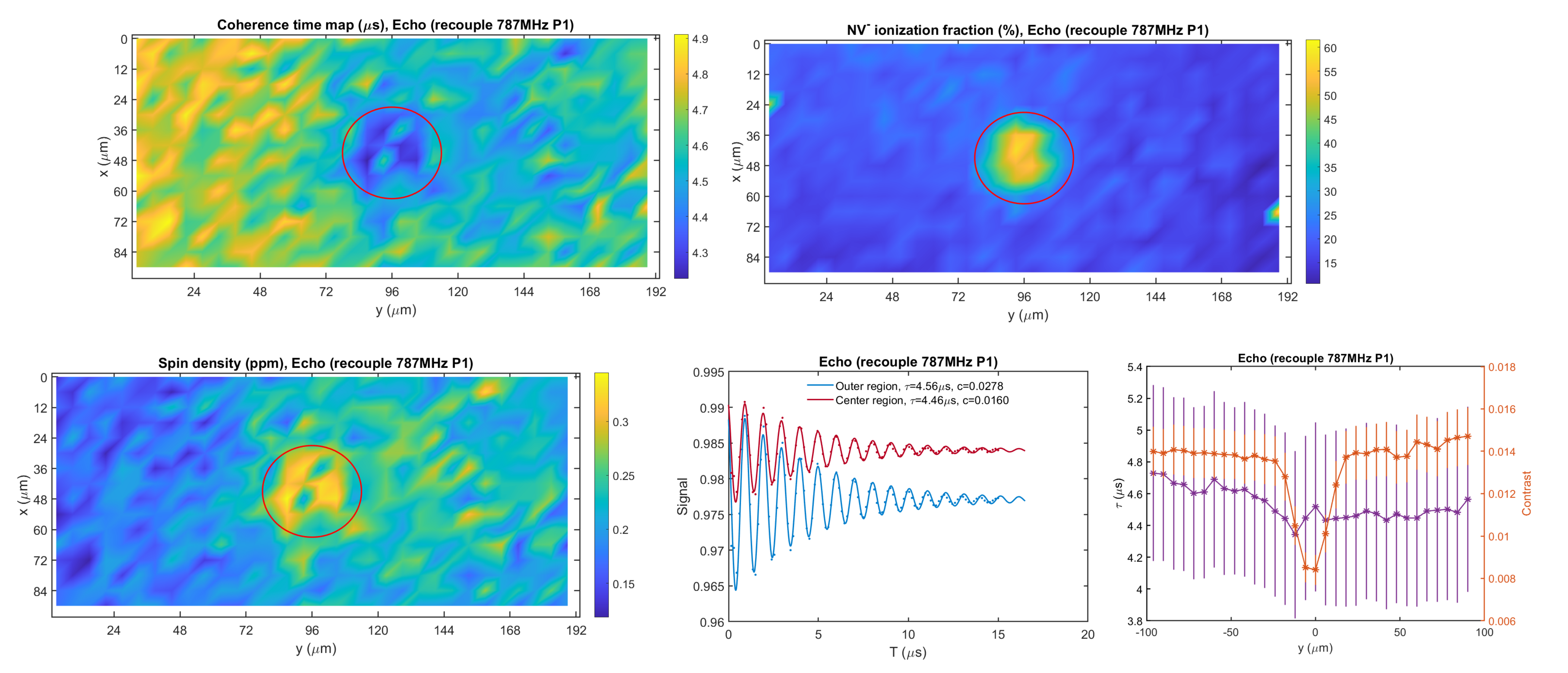}
\caption{\label{SuppFig_4} \textbf{DEER time evolution experiment with 787~MHz P1 recoupled.}
}
\end{figure*}

\begin{figure*}[htbp]
\centering \includegraphics[width=\textwidth]{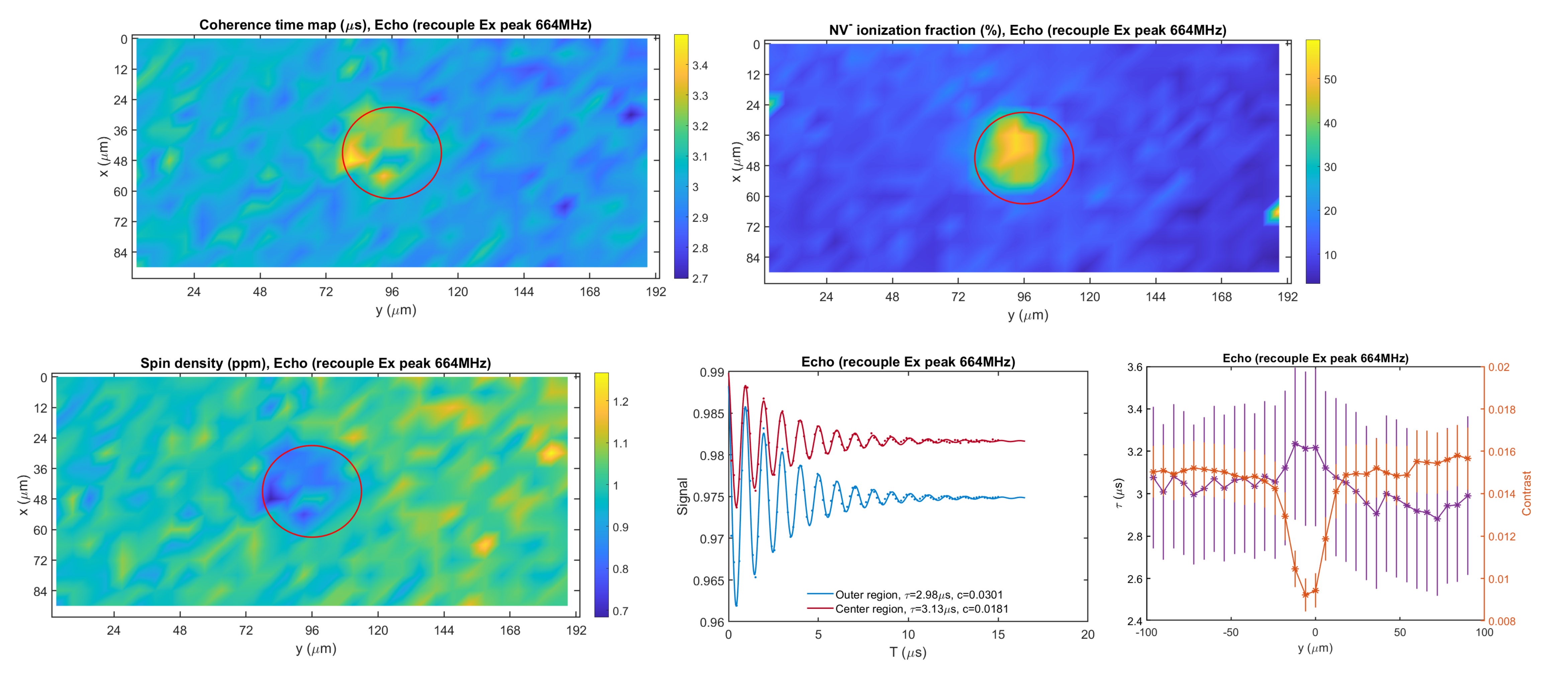}
\caption{\label{SuppFig_5} \textbf{DEER time evolution experiment with 664~MHz X recoupled.}
}
\end{figure*}

Figures~\ref{SuppFig_1},\ref{SuppFig_2},\ref{SuppFig_3},\ref{SuppFig_4},\ref{SuppFig_5} show the experimental data for the time evolution measurements while applying DEER sequence to recouple different resonance frequencies. Figure~\ref{SuppFig_6} shows the Rabi oscillation measurements of P1 bath spins at different resonance frequencies. Due to the imperfect sinusoidal shape of the oscillations, the data fitting is not perfect and the density plots are not clear. 

Since the data analysis for these figures are similar, we use one example (Fig.~\ref{SuppFig_1}) to explain the details in the data processing. The experimental sequence for the data here is shown in the main text figure. The coherence time map shows the NV coherence time $\tau$ extracted by fitting the time evolution signal to $S(T)=c_0+\frac{c}{2}\cos(\delta\omega t+\phi)e^{-t/\tau}$. In experiment, we set $\delta \omega=(2\pi)10$~MHz to visualize the time evolution of the spin echo. The NV$^-$ ionization fraction is obtained by calculating $c/c_{ref}$, where $c_{ref}$ is separately measured by an NV spin echo experiment under the same experimental condition without applying the pump laser beam and the bath recoupling pulse. The spin density of the P1 is extracted from Eq.~\eqref{Supp_eq:Ts*} by comparing the coherence time with the reference spin echo experiment $T_2$ under the same experimental condition without applying the pump laser beam and the bath recoupling pulse, $1/T_s^*=1/\tau-1/T_2$. The total signals in the regions inside and outside the circle are plotted in red and blue points respectively (the curves are their fittings). The coherence time $\tau$ and signal contrast $c$ obtained in the y-direction grouped pixels' signal (the same as the y-direction grouping) are also plotted as purple and orange curves, respectively.

Figure~\ref{SuppFig_2_2} shows the data for the time evolution measurements with a longer pump laser beam duration, while Figure~\ref{SuppFig_2_3} shows the reference data without the pump beam, where we see the charge manipulation effects due to the broad beam only.

\subsection{Details on the first-principles calculations}
We implemented the first-principles calculations to the NVH$^0$ and VH$_2^0$ defects in diamond. The defect structure is simulated by the supercell method~\cite{nieminen2007supercell} using a $3\times 3\times 3$ cubic supercell (with 216 atoms). The calculation used density functional theory (DFT) and projector-augmented wave (PAW) method carried out through Vienna $ab-initio$ simulation package (VASP)~\cite{kresse1996efficient,kresse1999ultrasoft}. Generalized gradient approximation is used for exchange-correlation interaction with the PBE functional~\cite{perdew1996generalized}. The cut-off energy is set as 400 eV, and a $3\times 3\times 3$ $k$-point mesh is sampled by Monkhorst-Pack scheme~\cite{monkhorst1976special}. The defect structure is fully relaxed with a residue force on each atom less than 0.01 eV/\AA , and the electronic energy converges to $10^{-7}$ eV. The zero-field splitting is then calculated from the DFT electronic ground state using~\cite{ivady2014pressure}:
\begin{equation}
D_{a b}=\mu_0 g_e^2 \beta^2 \sum_{i<j}^{p+q} \chi_{i j}\left\langle\Psi_{i j}\left(\mathbf{r}_1, \mathbf{r}_2\right)\left|\frac{r^2 \delta_{a b}-3 r_a r_b}{r^5}\right| \Psi_{i j}\left(\mathbf{r}_1, \mathbf{r}_2\right)\right\rangle.
\end{equation}
The calculated $D_{ab}$ of NVH$^0$ is:
\begin{equation}
D_{a b}(NVH^0)=\left[\begin{matrix}-0.76 &0 &0 \\ 0& -0.76& 0 \\ 0& 0& 1.53\end{matrix}\right] \text{GHz},
\end{equation}
where the $z$-axis is set as the $C_{3v}$ axis of NVH$^0$. The calculated $D_{ab}$ of VH$_2^0$ in the cubic axis (the axis is along the three lattice vectors of the cubic supercell of diamond, and we make the two hydrogen atoms in $xz$-plane) is:
\begin{equation}
D_{a b}(VH_2^0)=\left[\begin{matrix} -0.40 &0 & 2.21 \\ 0& 0.79& 0 \\ 2.21 & 0& -0.40\end{matrix}\right] \text{GHz},
\end{equation}

\begin{figure*}[htbp]
\centering \includegraphics[width=\textwidth]{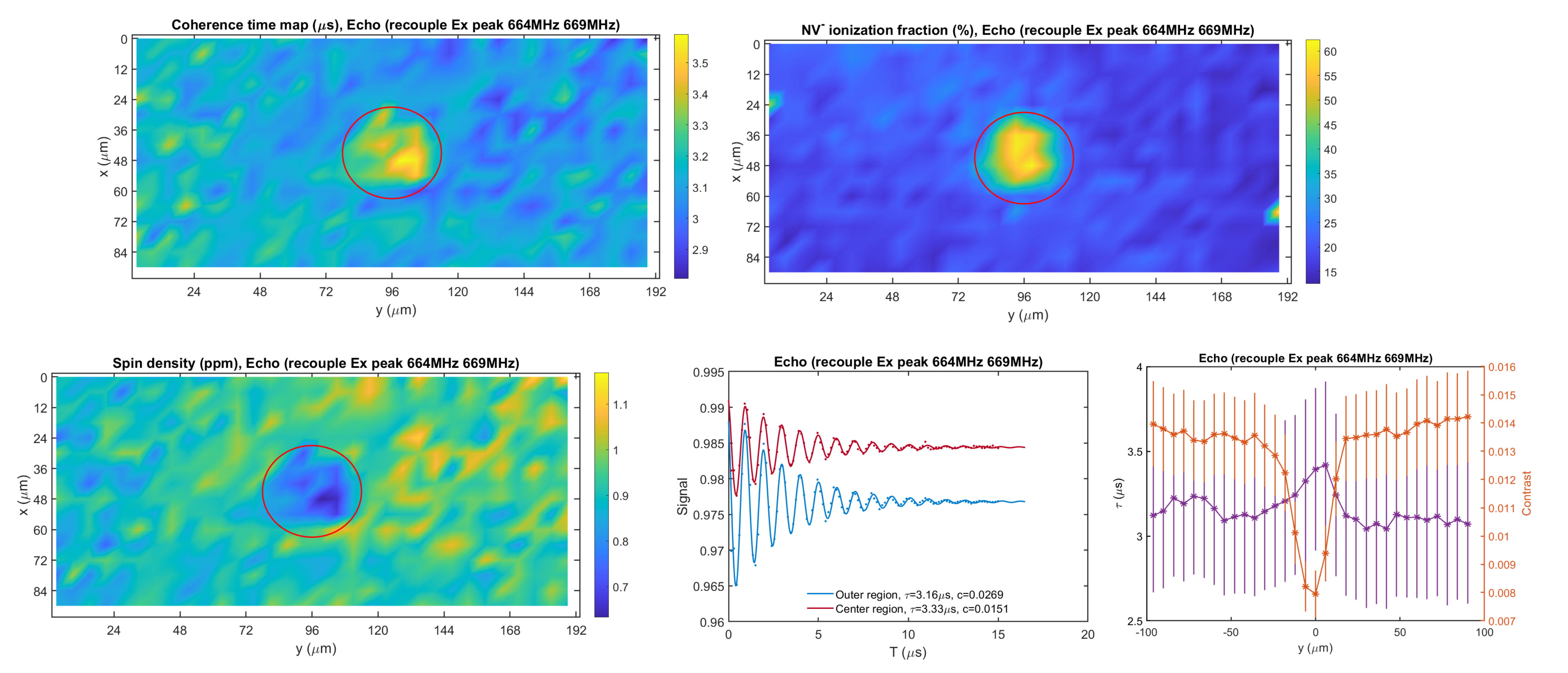}
\caption{\label{SuppFig_5_2} \textbf{DEER time evolution experiment with 664~MHz and 669~MHz X recoupled.} 
}
\end{figure*}

\begin{figure*}[htbp]
\centering \includegraphics[width=\textwidth]{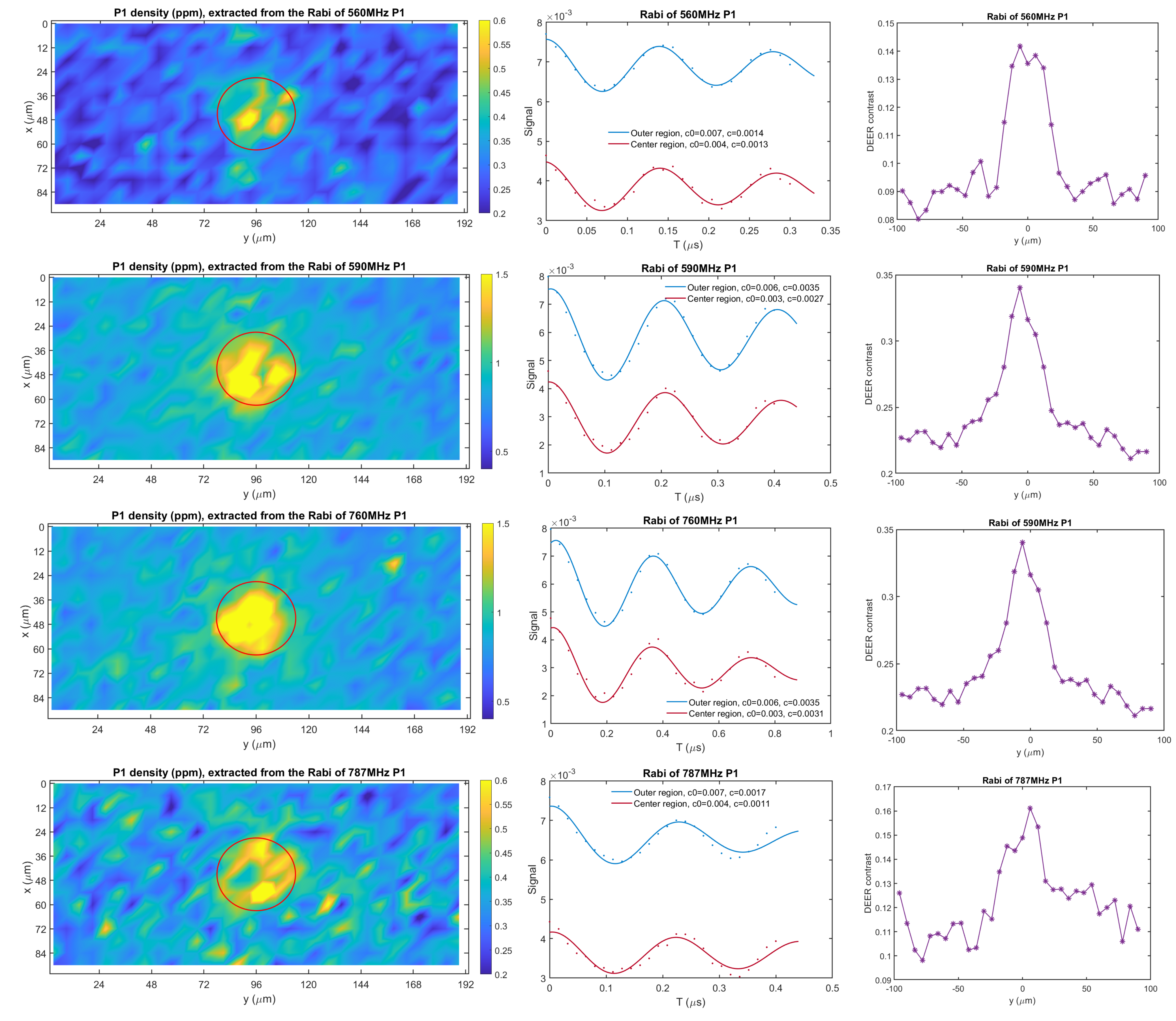}
\caption{\label{SuppFig_6} \textbf{Rabi of P1 centers.}
}
\end{figure*}

\begin{figure*}[htbp]
\centering \includegraphics[width=\textwidth]{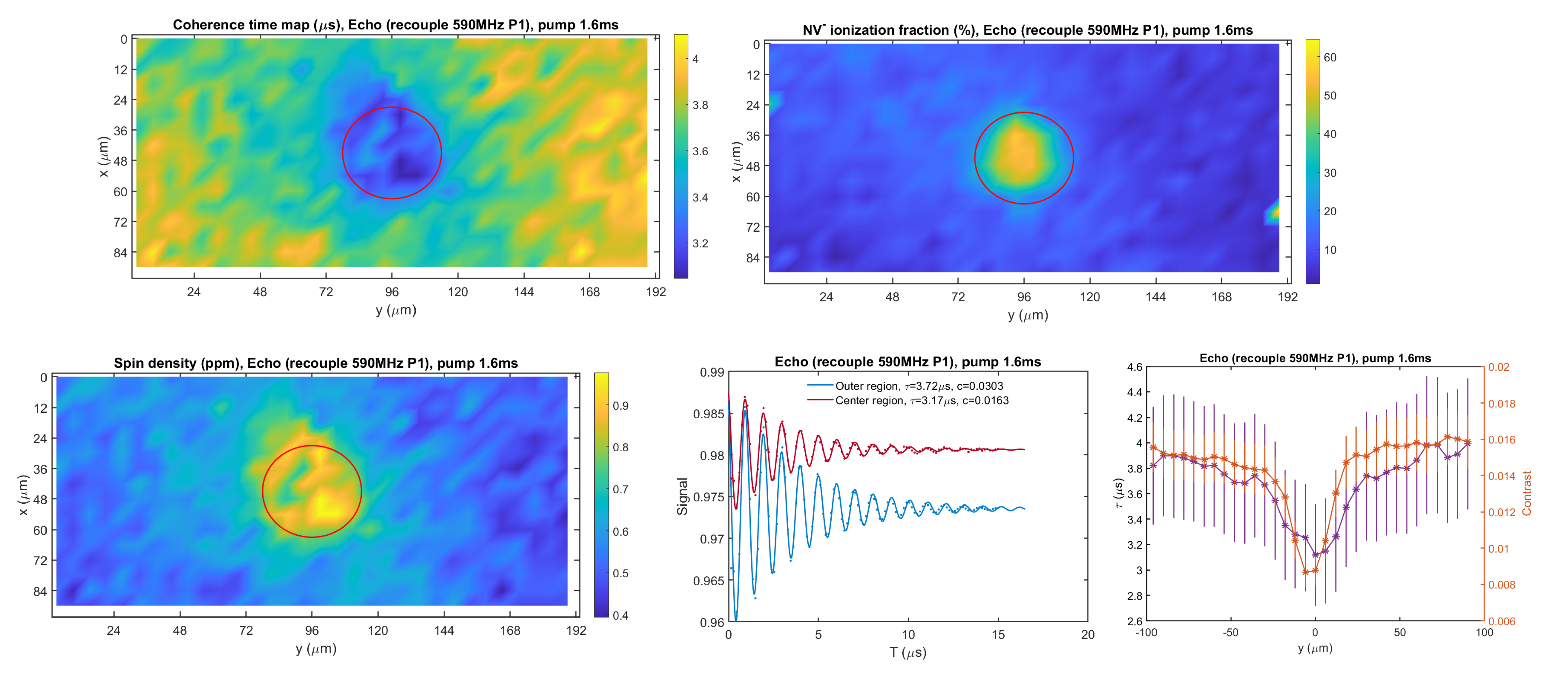}
\caption{\label{SuppFig_2_2} \textbf{DEER time evolution experiment with 590~MHz P1 recoupled.} The duration of the pump laser pulse is 1.6ms while other conditions are keeping the same as the experiment in Fig.~\ref{SuppFig_2}. 
}
\end{figure*}
\begin{figure*}[htbp]
\centering \includegraphics[width=\textwidth]{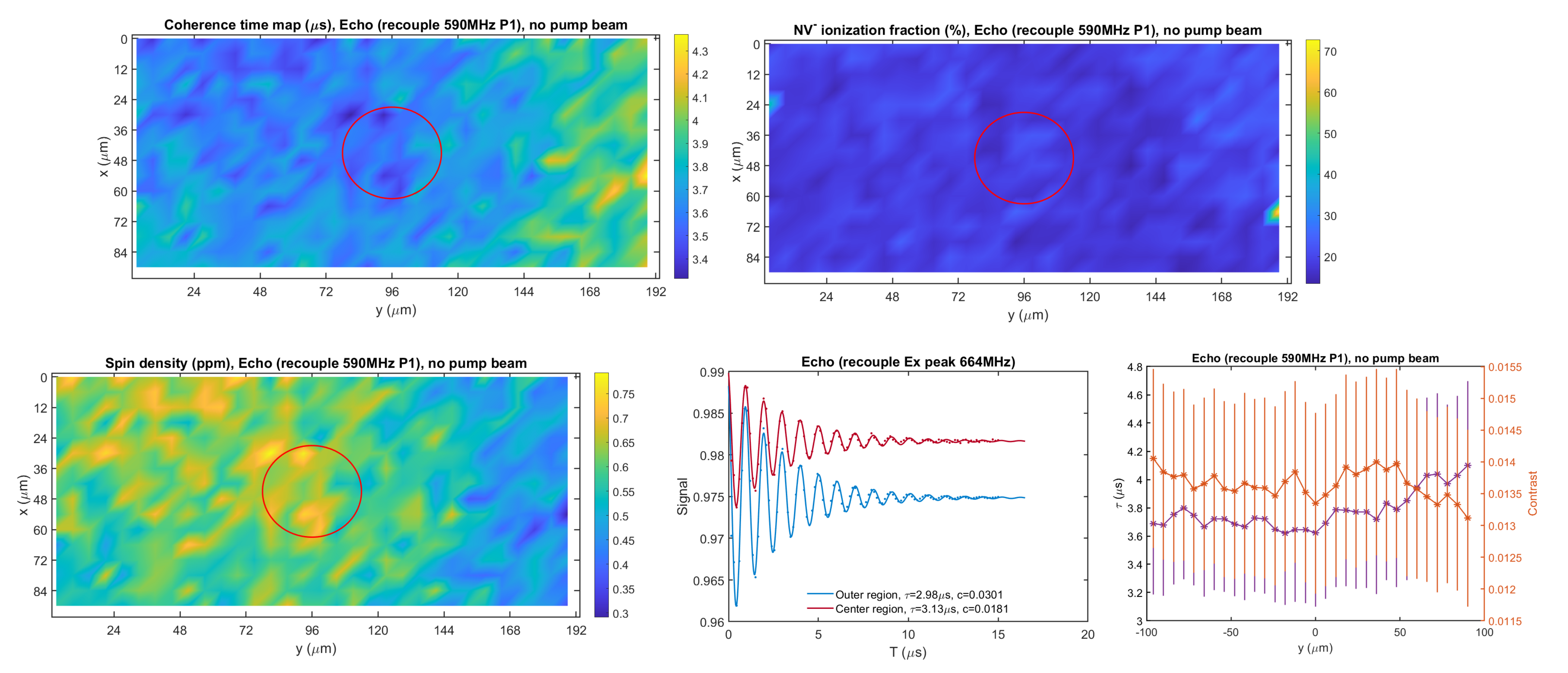}
\caption{\label{SuppFig_2_3} \textbf{DEER time evolution experiment with 590~MHz P1 recoupled.} No pump laser is applied while other experimental conditions are keeping the same as the experiment in Fig.~\ref{SuppFig_2}.
}
\end{figure*}


\clearpage


\end{document}